\documentclass[aps,preprintnumbers,nofootinbib,prx,superscriptaddress,tightenlines,notitlepage,twocolumn,showpacs,floatfix]{revtex4-1}
\usepackage{tabularx}
\usepackage{graphicx}
\usepackage{amssymb,bm,tensor}
\usepackage{textcomp}
\usepackage{xcolor}
\usepackage{amsmath}
\usepackage[varg]{txfonts}
\usepackage{enumerate}
\usepackage{mathtools}
\usepackage[normalem]{ulem}
\usepackage{bm} % bold math
\usepackage[OT1]{fontenc}
\usepackage[colorlinks]{hyperref}

\newcommand{\KIAA}{\affiliation{Kavli Institute for Astronomy and Astrophysics,
Peking University, Beijing 100871, China}}

\begin{document}

\title[Search for anisotropic birefringence in GWTC-1]{Combined search for
anisotropic birefringence in the gravitational-wave transient catalog
GWTC-1}
\date{\today}
\author{Lijing Shao}\email{lshao@pku.edu.cn}\KIAA

\begin{abstract} 
The discovery of gravitational waves (GWs) provides an unprecedented arena
to test general relativity, including the gravitational Lorentz invariance
violation (gLIV). In the propagation of GWs, a generic gLIV leads to
anisotropy, dispersion, and birefringence. GW events constrain the
anisotropic birefringence particularly well. Kosteleck\'y and Mewes (2016)
performed a preliminary analysis for GW150914. We improve their method and
extend the analysis {\it systematically} to the whole GW transient catalog,
GWTC-1. This is the first global analysis of the spacetime anisotropic
Lorentzian structure with a catalog of GWs, where multiple events are
crucial in breaking the degeneracy among gLIV parameters. With the absence
of abnormal propagation, we obtain new limits on 34 coefficients for gLIV
in the nonminimal gravity that surpass previous limits by $\sim
10^2$--$10^5$.
\end{abstract}

\maketitle

\section{Introduction}
\label{sec:intro}
%---------------------------------------------------------------------

Lorentz invariance is a cherished symmetry laying to the heart of modern
physics. However, motivated by contemporary open questions, there are {\it
good} reasons to believe that, the Lorentz symmetry breaks at some yet
unknown energy scale~\cite{Jacobson:2005bg, Tasson:2014dfa}. For example,
in the string theory it is perceived to have a Higgs-like spontaneous
symmetry breaking of the Lorentz invariance~\cite{Kostelecky:1988zi,
Kostelecky:1991ak} which leads to various observable
phenomena~\cite{Mattingly:2005re, Kostelecky:2008ts}. While the Lorentz
invariance violation (LIV) in principle happens in different species
sectors~\cite{Carroll:1989vb, Colladay:1996iz, Colladay:1998fq}, it is the
most interesting to study the {\it gravitational} LIV (gLIV) due to the
following fact~\cite{Kostelecky:2003fs, Hees:2016lyw, Shao:2016ezh,
Tasson:2016xib}. Up to now, the canonical theory of gravitation, namely the
general relativity, while being extremely faithful in describing various
gravity experiments~\cite{Will:2014kxa, Berti:2015itd}, still refuses to
embed into the framework of quantum field theory which successfully
describes the other three fundamental interactions in a unified way.
Therefore, searches for gLIV are closely linked to searches for
quantum-gravity candidate theories~\cite{AmelinoCamelia:1997gz,
Gambini:1998it, AmelinoCamelia:2008qg, Berti:2015itd}.

Nowadays the most popular framework to investigate the possibility of gLIV
is the standard-model extension (SME)~\cite{Kostelecky:2003fs} and the
parameterized post-Newtonian formalism~\cite{Will:2014kxa, Will:2018bme}.
We will focus on the former. The SME is an agnostic effective field theory
which incorporates all gauge-preserving, Lorentz-violating,
energy-momentum-conserving operators that are constructed from the field
operators in general relativity and particle physics. New Lorentz-covariant
operators are built from contraction of known fields with LIV coefficients.
Unless being protected, operators with lower mass dimension are expected to
dominate the observables as {\it per} the spirit of effective field
theories. Here we focus on the subset of the SME that deals with the pure
gravity sector; interested readers are referred to
Ref.~\cite{Kostelecky:2010ze} for discussions on the matter-gravity
couplings. In an effective-field-theoretic framework, to be compatible with
the Riemann-Cartan geometry, the symmetry breaking shall be {\it
spontaneous} instead of {\it explicit}~\cite{Bluhm:2004ep}. Therefore the
SME, while being particle Lorentz-violating, is observer Lorentz-invariant.

The convenience in using the SME to test the Lorentz symmetry has resulted
in a flourish of studies during the past decades from different kinds of
experiments concerning various particle species~\cite{Kostelecky:2008ts}.
For the gravity sector, constraints come from lunar laser
ranging~\cite{Battat:2007uh, Bourgoin:2017fpo}, atom
interferometers~\cite{Muller:2007es}, cosmic
rays~\cite{Kostelecky:2015dpa}, precision pulsar timing~\cite{Shao:2014oha,
Shao:2014bfa, Jennings:2015vma, Shao:2018vul, Shao:2019cyt, Shao:2019tle},
planetary orbital dynamics~\cite{Hees:2015mga}, laboratory short-range
experiments~\cite{Bailey:2014bta, Shao:2016cjk, Kostelecky:2016uex,
Shao:2018lsx}, super-conducting gravimeters~\cite{Flowers:2016ctv}, and
recently gravitational waves (GWs)~\cite{Kostelecky:2016kfm, Mewes:2019dhj,
Xu:2019gua}; see \citet{Hees:2016lyw} for a comprehensive review. These
constraints are complementary to each other, and in many cases they are
individually competent at probing different parts of the parameter space.
For example, timing of binary pulsars was shown to be good at
systematically probing the lowest-order CPT-violating~\cite{Bailey:2017lbo,
Shao:2018vul}, as well as gravitational
weak-equivalence-principle-violating~\cite{Bailey:2016ezm, Shao:2019cyt},
operators via the post-Newtonian dynamics, while laboratory short-range
experiments excel in constraining the nonminimal operators in a static
Newtonian setup~\cite{Kostelecky:2016uex}.

A newly emerging probe to gLIV phenomena in gravity is the recently
discovered GWs by the Advanced LIGO/Virgo detectors~\cite{Abbott:2016blz,
Kostelecky:2016kfm, LIGOScientific:2018mvr}. The propagation of GWs has
become one of the central topics in looking for gLIV
clues~\cite{Will:1997bb, Mirshekari:2011yq, Yunes:2016jcc, Abbott:2017vtc,
Nishizawa:2017nef, Arai:2017hxj, Wang:2017igw, Qiao:2019wsh, Zhao:2019xmm,
Wang:2020pgu}. The existence of gLIV would generally cause anisotropy,
dispersion, and birefringence~\cite{Kostelecky:2016kfm, Mewes:2019dhj}.
Following the theoretical work by \citet{Will:1997bb} and
\citet{Mirshekari:2011yq}, the LIGO/Virgo Collaboration have conducted
extensive tests of the GW dispersion relation caused by the gLIV or a
massive graviton~\cite{Abbott:2017vtc, LIGOScientific:2019fpa}. However,
most of the past study focused on the boost breaking, and ignored the
breaking of the rotational symmetry. While the commutator of two boost
generators gives a rotation generator, it is {\it inevitable} to host the
rotation breaking when the boost symmetry is broken, unless the object
under investigation is exactly at rest with respect to a preferred frame.
Such a preferred frame may not even exist in effective field
theories~\cite{K.Nordtvedt:1976zz, Bailey:2006fd}. This motivates us to
look at the anisotropic gLIV and test it thereof in a more generic way.

In this work we follow the preliminary analysis done by
\citet{Kostelecky:2016kfm}, and extend it {\it systematically} with all the
GW events detected in the first and second observing
runs~\cite{LIGOScientific:2018mvr}. It is the first study of this kind, and
probes the gLIV with a {\it globally coherent} approach. Using the fact
that the SME is {\it observer} Lorentz-invariant, we combine different GW
events from the GW transient catalog, GWTC-1~\cite{LIGOScientific:2018mvr},
including ten binary black holes (BBHs) and one binary neutron star (BNS),
GW\,170817~\cite{TheLIGOScientific:2017qsa}; see Appendix~\ref{app:gwtc}
for basic parameters of GW events in the GWTC-1. A global analysis with the
basis of spin-weighted spherical harmonics provides us new limits on the
gLIV coefficients in the nonminimal gravity at mass dimensions 5 and 6.
These limits are orders of magnitude tighter than the existing ones. GW
observations truly represent precious treasures in studying various gravity
theories. While more detections are continuously being made, the
investigation in this paper will be improved dramatically and new clues to
quantum gravity might be drawn.

The paper is organized as follows. In the next section, we briefly discuss
the modified dispersion relation for the linearized gravity in SME, and its
effect on the cosmological GW propagation. Anisotropic birefringent effects
can be constrained particularly well with GWs. Therefore we focus on the
anisotropic birefringence in Sec.~\ref{sec:bire}, and lay down the
strategies for the practical application to the \mbox{GWTC-1}. In
Sec.~\ref{sec:res} we present detailed Monte Carlo analysis using the
posteriors on the GW parameters for the 11 events in the catalog,
provided by the LIGO/Virgo Collaboration. Numerical constraints on 34 gLIV
coefficients are obtained with the {\it
maximal-reach}~\cite{Tasson:2019kuw} and {\it global} approaches. The
global analysis, simultaneously with tens of gLIV parameters analyzed, is
done for the first time with GWs in a coherent way. It greatly extends
previous work done by the LIGO/Virgo Collaboration~\cite{Abbott:2017vtc,
Abbott:2018lct, LIGOScientific:2019fpa} and \citet{Kostelecky:2016kfm}.
Finally, Sec.~\ref{sec:diss} summarizes the paper and discusses future
directions for improvement. For readers' convenience, extra information on
the GWTC-1 catalog, the spin-weighted spherical harmonics, and a fitting
formula to the GW peak frequency are provided respectively in
Appendices~\ref{app:gwtc}, \ref{app:spin:harmonics}, and
\ref{app:peak:frequency}.

Throughout the paper, we follow conventions used by \citet{Misner:1974qy}
and \citet{Kostelecky:2003fs}. Unless otherwise stated, we use natural
units where $\hbar=G=c=1$.

%---------------------------------------------------------------------
\section{Dispersion and propagation of gravitational waves}
\label{sec:dispersion}
%---------------------------------------------------------------------

The gravity sector in the SME was given fully in the Riemann-Cartan
spacetime~\cite{Kostelecky:2003fs}. However, we restrict ourselves to a
4-dimensional Riemannian spacetime, and only consider the part of spacetime
where, after fixing the gauge, the linearized gravity is a sufficiently
good approximation. The metric, $g_{\mu\nu} = \eta_{\mu\nu} + h_{\mu\nu}$,
is expanded around the Minkowski metric, $\eta_{\mu\nu}$, with a
perturbation, $h_{\mu\nu}$. \citet{Kostelecky:2016kfm} constructed the
general quadratic Lagrangian density for GWs in the presence of gLIV
operators of arbitrary mass dimension $d$,
%--
\begin{equation}\label{eq:LhatK}
  {\cal L}_{ {\cal K}^{(d)}} = \frac{1}{4} h_{\mu\nu} \hat{\cal
  K}^{(d)\mu\nu\rho\sigma} h_{\rho\sigma} \,,
\end{equation}
%--
where $\hat{\cal K}^{(d)\mu\nu\rho\sigma} = {\cal K}^{(d)\mu\nu\rho\sigma
i_1 i_2 \cdots i_{d-2}} \partial_{i_1} \partial_{i_2} \cdots
\partial_{i_{d-2}}$ with the ``hat'' denoting its operator nature, and
${\cal K}^{(d)\mu\nu\rho\sigma i_1 i_2 \cdots i_{d-2}} $ are normal
tensorial components whose mass dimension is $4-d$.

With the help of the Young tableaux, \citet{Kostelecky:2016kfm,
Kostelecky:2017zob} studied the irreducible decomposition of the operator
$\hat{\cal K}^{(d)\mu\nu\rho\sigma}$ thoroughly. They found that three
classes of operators are invariant under the infinitesimal gauge
transformation $h_{\mu \nu} \rightarrow h_{\mu \nu}+\partial_{\mu}
\xi_{\nu}+\partial_{\nu} \xi_{\mu}$. Imposing such a gauge symmetry, one
arrives at a {\it complete} gauge-invariant quadratic Lagrangian density,
%--
\begin{equation}\label{eq:L}
  {\cal L} = {\cal L}_0 + \frac{1}{4} h_{\mu\nu} \left(
  \hat{s}^{\mu\rho\nu\sigma} + \hat{q}^{\mu\rho\nu\sigma} +
  \hat{k}^{\mu\nu\rho\sigma} \right) h_{\rho\sigma} \,,
\end{equation}
%--
where $ {\cal L}_0 = \frac{1}{4} \epsilon^{\mu\rho\alpha\kappa}
\epsilon^{\nu\sigma\beta\lambda} \eta_{\kappa\lambda} h_{\mu\nu}
\partial_\alpha \partial_\beta h_{\rho\sigma}$ is the linearized
Einstein-Hilbert Lagrangian. Operators $\hat{s}^{\mu\rho\nu\sigma}$,
$\hat{q}^{\mu\rho\nu\sigma}$, and $\hat{k}^{\mu\nu\rho\sigma}$ are the sums
over $d$ of each of three irreducible sets; see Table~1 and Eq.~(3) in
Ref.~\cite{Kostelecky:2016kfm} for their relations to $\hat{\cal
K}^{(d)\mu\nu\rho\sigma}$. These operators have the following symmetries:
%--
\begin{enumerate}[(i)]
  \item $\hat{s}^{\mu\rho\nu\sigma}$ is anti-symmetric in both
  ``$\mu\rho$'' and ``$\nu\sigma$'';
  \item $\hat{q}^{\mu\rho\nu\sigma}$ is anti-symmetric in ``$\mu\rho$'' and
  symmetric in ``$\nu\sigma$'';
  \item $\hat{k}^{\mu\nu\rho\sigma}$ is totally symmetric.
\end{enumerate}
%--  
In addition, any contraction of these operators with a derivative is zero.

Similar to the previous study in the photon sector of the electrodynamic
SME~\cite{Kostelecky:2002hh, Kostelecky:2009zp}, the covariant dispersion
relation for the two tensor modes of a GW with 4-momentum $p^\mu = \left(
\omega, \bm{p} \right)$ is~\cite{Kostelecky:2016kfm},
%--
\begin{equation}\label{eq:DR}
  \omega = \left( 1 - \zeta^0 \pm \sqrt{\left( \zeta^1 \right)^2 + \left(
  \zeta^2 \right)^2 + \left( \zeta^3 \right)^2} \right) p \,,
\end{equation}
%--
where
%--
\begin{equation}
  \zeta^0 = \frac{1}{4p^2} \left( - \tensor{\hat s}{^{\mu\nu}_{\mu\nu}} +
  \frac{1}{2} \tensor{\hat k}{^{\mu\nu}_{\mu\nu}} \right) \,,
\end{equation}
%--
\begin{equation}
  \left( \zeta^1 \right)^2 + \left( \zeta^2 \right)^2 = \frac{1}{8p^4}
  \left( \hat{k}^{\mu\nu\rho\sigma} \hat{k}_{\mu\nu\rho\sigma} -
  \tensor{\hat k}{^{\mu\rho}_{\nu\rho}} \tensor{\hat
  k}{_{\mu\sigma}^{\nu\sigma}} + \frac{1}{8} \tensor{\hat
  k}{^{\mu\nu}_{\mu\nu}} \tensor{\hat k}{^{\rho\sigma}_{\rho\sigma}}
  \right) \,,
\end{equation}
%--
\begin{equation}
  \left( \zeta^3 \right)^2 = \frac{1}{16 p^4} \left[\left( \tensor{\hat
  q}{^{\mu\rho\nu}_\rho} + \tensor{\hat q}{^{\nu\rho\mu}_\rho} \right)
  \tensor{\hat q}{_{\mu\sigma\nu}^\sigma} -\frac{1}{2}
  \hat{q}^{\mu\rho\nu\sigma} \hat{q}_{\mu\rho\nu\sigma} -
  \hat{q}^{\mu\nu\rho\sigma} \hat{q}_{\mu\rho\nu\sigma} \right] \,.
\end{equation}
%--
Here the decomposition is done by the handedness of GWs, instead of the
usual ``$+$'' and ``$\times$'' modes. In above equations, $\zeta^0$ and
$\zeta^3$ are rotation scalars, $\zeta^1$ and $\zeta^2$ are helicity-4
tensors, and the derivative ``$\partial_\mu$'' in operators is understood
to be replaced by ``$ip_\mu$''~\cite{Kostelecky:2016kfm}.

%---------------------------------------------------------------------
\begin{figure*}[t!]
  \centering
  \includegraphics[width=16cm]{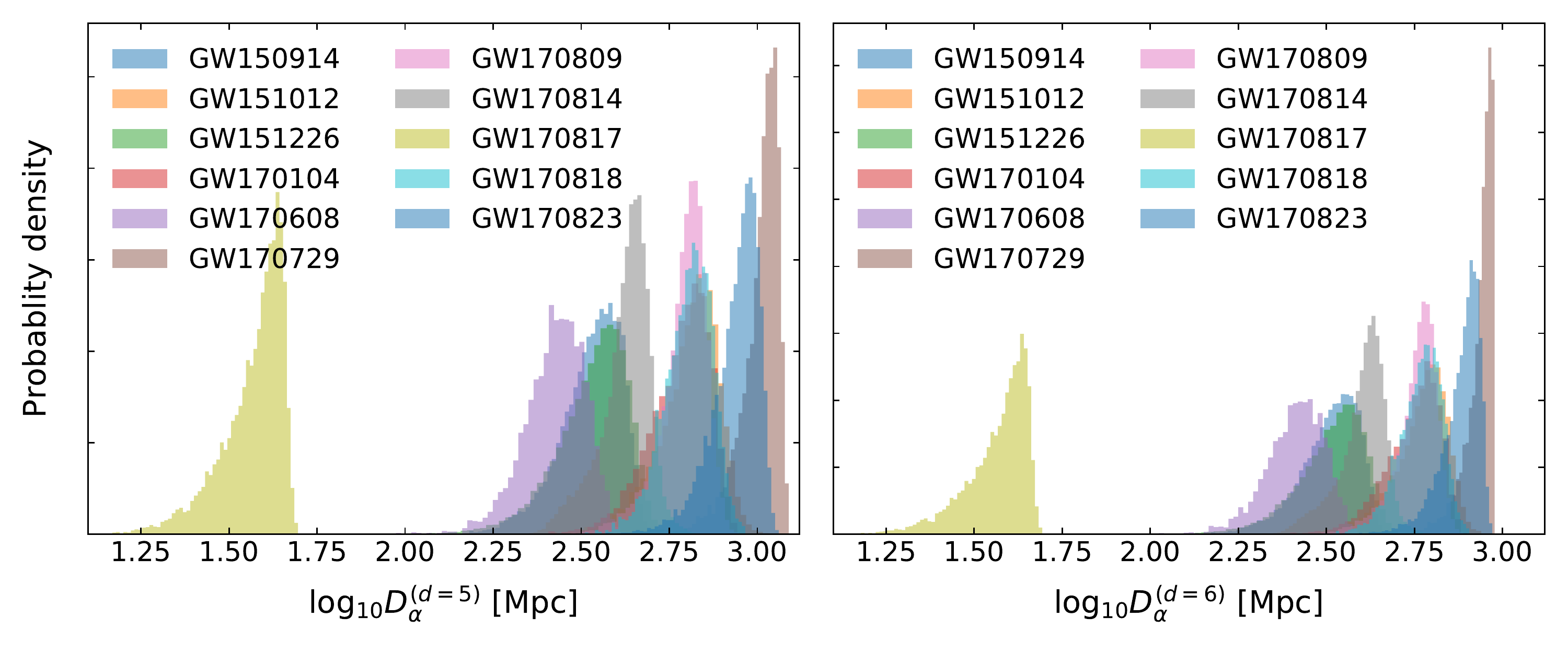}
\caption{Probability density for $D_\alpha^{(d=5)}$ and $D_\alpha^{(d=6)}$, 
derived using the posterior samples provided by the
LIGO/Virgo Collaboration~\cite{LIGOScientific:2018mvr, GWTC1:catalog, GWTC1:PE}. 
\label{fig:Da}}
\end{figure*}
%---------------------------------------------------------------------

To report experimental constraints on the coefficients for Lorentz/CPT
violation in the SME, it is conventional to use the canonical Sun-centered
celestial-equatorial frame~\cite{Kostelecky:2002hh}. Concerning the
rotational behavior, it is useful to decompose $\zeta^\alpha$ with
spin-weighted spherical harmonics~\cite{Kostelecky:2009zp,
Kostelecky:2016kfm},
%--
\begin{eqnarray}
  \zeta^0 &=& \sum_{djm} \omega^{d-4} Y_{jm}\left( \hat{\bm{n}} \right)
  k^{(d)}_{(I)jm} \,, \label{eq:kI}\\
  %--
  \zeta^1 \mp i\zeta^2 &=& \sum_{djm} \omega^{d-4} {}_{\pm 4} Y_{jm}\left(
  \hat{\bm{n}} \right) \left[ k^{(d)}_{(E)jm} \pm i k^{(d)}_{(B)jm} \right]
  \,, \label{eq:kEB}\\
  %--
  \zeta^3 &=& \sum_{djm} \omega^{d-4} Y_{jm}\left( \hat{\bm{n}} \right)
  k^{(d)}_{(V)jm} \,, \label{eq:kV}
\end{eqnarray}
%--
where $\hat{\bm{n}} \equiv - \hat{\bm{p}}$ is the direction to the source,
and $|s| \leq j \leq d-2$; see Appendix~\ref{app:spin:harmonics} for more
details on the spin-weighted spherical harmonics. It was shown
that~\cite{Kostelecky:2016kfm},
%--
\begin{enumerate}[(i)]
  \item anisotropic effects are governed by the coefficients with $j \neq
  0$;
  \item frequency-dependent dispersions are governed by all coefficients
  except $k^{(4)}_{(I)jm}$;
  \item {\it vacuum birefringent} effects are governed by all coefficients
  except $k^{(d)}_{(I)jm}$.
\end{enumerate}
%--
Besides, $d\geq 4$ is even for $k^{(d)}_{(I)jm}$; $d \geq 5$ is odd for
$k^{(d)}_{(V)jm}$; and $d \geq 6$ is even for $k^{(d)}_{(E)jm}$ and
$k^{(d)}_{(B)jm}$.

Assuming that the corrections to Einstein's general relativity are small,
we obtain the {\it phase} speed of GWs from Eq.~(\ref{eq:DR}), $ v_{\rm GW} = \omega/p =  1
- \zeta^0 \pm \sqrt{ \sum_{i=1}^3 \left(\zeta^i\right)^2} $. In accordance
to the spirit of effective field theories, we further assume that the gLIV
happens {\it dominantly} at a specific mass dimension $d$. If it had
happened at multiple dimensions, we equivalently take the dimension $d$,
usually the lowest relevant mass dimension where gLIV happens, which
introduces the maximum effect. By this assumption we ease the summation
over ``$d$'' in Eqs.~(\ref{eq:kI}--\ref{eq:kV}), and denote the original
$\zeta^\alpha$ as $\zeta^\alpha_{(d)}$ ($\alpha=0,1,2,3$). Further, we
introduce a notation, $\tilde{\zeta}^\alpha_{(d)} \left( \hat{\bm{n}}
\right) \equiv \zeta^\alpha_{(d)} \left( \hat{\bm{n}} \right) \times
\omega^{4-d}$ which are {\it energy-independent}; for example,
$\tilde{\zeta}^0_{(d)} \left( \hat{\bm{n}} \right) = \sum_{jm} Y_{jm}
\left( \hat{\bm{n}} \right) k^{(d)}_{(I)jm}$. With these considerations,
the speed of GWs simplifies to,
%--
\begin{equation}
    v_{\rm GW}^{(d)} = 1 - \omega^{d-4} \left\{ \tilde\zeta^0_{(d)} \mp
  \sqrt{\sum_{i=1}^3\left( \tilde\zeta^i_{(d)} \right)^2} \right\} 
  \,,
\end{equation}
%--
where an important fact is that the terms inside the curly bracket is,
while being direction-dependent, energy-independent.

%---------------------------------------------------------------------
\def\arraystretch{1.5}
\begin{table}[t]
\caption{In the table we show (i) the GW peak frequency $f_{\rm GW}$ at the 
merger using the fit in Eq.~(\ref{eq:fGW:peak}), and (ii) the distance-like
quantity, $D_\alpha^{(d)}$, defined in Eq.~(\ref{eq:Dalpha}), for mass
dimensions $d=5$ and $d=6$. The uncertainties are given at the $90\%$
confidence level. Notice that, because the merger part for the BNS was not
observed~\cite{TheLIGOScientific:2017qsa,Abbott:2018wiz}, in our test we
{\it conservatively} use $f_{\rm GW} = 800$\,Hz for GW170817. 
\label{tab:gw:D:f}}
\centering
\begin{tabular}{p{2cm}p{1.6cm}p{2.2cm}p{2.2cm}}
\hline\hline
& $f_{\rm GW}$ [Hz] & $D_\alpha^{(d=5)}$ [Mpc] & $D_\alpha^{(d=6)}$ [Mpc] \\
\hline
GW150914 & $174 _{ -7 } ^{+ 6 }$ & $350 _{ -120 } ^{+ 90 }$ & $330 _{ -110 } ^{+ 80 }$ \\
GW151012 & $307 _{ -68 } ^{+ 35 }$ & $660 _{ -220 } ^{+ 170 }$ & $610 _{ -190 } ^{+ 140 }$ \\
GW151226 & $547 _{ -122 } ^{+ 45 }$ & $350 _{ -130 } ^{+ 110 }$ & $340 _{ -120 } ^{+ 100 }$ \\
GW170104 & $220 _{ -22 } ^{+ 18 }$ & $620 _{ -200 } ^{+ 150 }$ & $570 _{ -180 } ^{+ 120 }$ \\
GW170608 & $619 _{ -97 } ^{+ 25 }$ & $270 _{ -80 } ^{+ 80 }$ & $260 _{ -80 } ^{+ 70 }$ \\
GW170729 & $146 _{ -21 } ^{+ 18 }$ & $1050 _{ -260 } ^{+ 120 }$ & $890 _{ -180 } ^{+ 60 }$ \\
GW170809 & $196 _{ -18 } ^{+ 12 }$ & $640 _{ -180 } ^{+ 110 }$ & $580 _{ -160 } ^{+ 90 }$ \\
GW170814 & $209 _{ -10 } ^{+ 7 }$ & $430 _{ -130 } ^{+ 90 }$ & $410 _{ -120 } ^{+ 70 }$ \\
GW170817 & $4243 _{ -81 } ^{+ 24 }$ & $39 _{ -14 } ^{+ 6 }$ & $38 _{ -14 } ^{+ 6 }$ \\
GW170818 & $180 _{ -14 } ^{+ 12 }$ & $650 _{ -170 } ^{+ 150 }$ & $590 _{ -140 } ^{+ 120 }$ \\
GW170823 & $169 _{ -20 } ^{+ 17 }$ & $900 _{ -250 } ^{+ 150 }$ & $790 _{ -190 } ^{+ 100 }$ \\
\hline
\end{tabular}
\end{table}
%---------------------------------------------------------------------

Now consider two gravitons emitted at $t_e$ and $t_e^\prime$ with energies
$\omega_e$ and $\omega^\prime_e$ respectively. After traveling over a same
{\it comoving} distance~\cite{Will:1997bb, Jacob:2008bw,
Mirshekari:2011yq}, they arrive at a GW detector on the Earth at $t_a$ and
$t_a^\prime$. Following the method developed by \citet{Will:1997bb} and
\citet{Mirshekari:2011yq}, we derive
%--
\begin{equation} \label{eq:Delta:ta}
  \Delta t_a = (1+z) \left\{ \Delta t_e + \Delta \omega_e^{d-4} D_\alpha^{(d)}
  \left[ \tilde\zeta^0_{(d)} \mp \sqrt{ \sum_{i=1}^3 \left( \tilde\zeta^i_{(d)}
\right)^2} \right] \right\} \,,
\end{equation}
%--
where $\Delta t_e \equiv t_e - t_e^\prime$, $\Delta t_a \equiv t_a -
t_a^\prime$, $\Delta \omega_e^{d-4} \equiv \omega_e^{d-4} -
{\omega_e^\prime}^{d-4}$, and $z$ is the cosmological redshift. A new
distance-like quantity in the above equation is defined
as~\cite{Will:1997bb},
%--
\begin{equation}
  D_\alpha^{(d)} \equiv \frac{\left( 1+z \right)^{3-d}}{H_0} \int_{0}^{z}
  \frac{\left( 1+ z^\prime \right)^{d-4}}{\sqrt{\Omega_m \left( 1+z^\prime
  \right)^3 + \Omega_\Lambda}} {\rm d} z^\prime \,, \label{eq:Dalpha}
\end{equation}
%--
where $H_0 = \left( 67.4 \pm 0.5 \right) \, {\rm km}\,{\rm s}^{-1}\,{\rm
Mpc}^{-1}$ is the Hubble constant, $\Omega_m = 0.315 \pm 0.007$ is the
fraction of matter energy density, and $\Omega_\Lambda = 0.685 \pm 0.007$
is the fraction of vacuum energy density in our current
Universe~\cite{Aghanim:2018eyx}. We have used the standard $\Lambda$CDM
model, which should be rather precise for the GW events with relatively low
redshifts. For the 11 GW events in the GWTC-1, their $D_\alpha^{(d)}$
for $d=5$ and $d=6$ are given in Fig.~\ref{fig:Da} and
Table~\ref{tab:gw:D:f}.

%---------------------------------------------------------------------
\section{Constraining anisotropic birefringence}
\label{sec:bire}
%---------------------------------------------------------------------

The modified dispersion relation (\ref{eq:DR}) introduces anisotropy,
dispersion, and birefringence to the propagation of
GWs~\cite{Kostelecky:2016kfm}. As the first application,
\citet{Kostelecky:2016kfm} used the observation that there is no indication
of {\it mode splitting} at the amplitude peak of
GW150914~\cite{Abbott:2016blz}. They took a rough value for the upper limit
of the time difference for the arrival of two circular modes, $\Delta t
\leq 3$\,ms, and used a central frequency $f \sim 100$\,Hz, to derive the
limit on the difference in the propagation speed between the two circularly
polarized CPT-conjugate eigenmodes. They obtained the first constraint on
the gLIV in the pure-gravity sector with $d=5$,
%--
\begin{align} \label{eq:KM:limit1}
  \left| \sum_{jm} Y_{jm}\left( \theta, \phi \right) k^{(5)}_{(V)jm}
  \right| \leq 2 \times 10^{-14} \, {\rm m} \,,
\end{align}
%--
and a competitive limit to existing laboratory bounds on birefringent
coefficients at $d=6$,
%--
\begin{align} \label{eq:KM:limit2}
  \left| \sum_{jm} {}_{\pm4} Y_{jm} \left( \theta,\phi \right) \left(
  k^{(6)}_{(E)jm} \pm i k^{(6)}_{(B)jm}\right) \right| \leq 8 \times
  10^{-9} \, {\rm m}^2 \,,
\end{align}
%--
for $\theta \simeq 160^\circ$ and $\phi \simeq
120^\circ$~\cite{Kostelecky:2016kfm}, where $(\theta, \phi)$ is the (very)
rough sky position of GW150914 in the Sun-centered celestial-equatorial
frame. The results are, though heuristic, very encouraging.

Equations~(\ref{eq:KM:limit1}) and (\ref{eq:KM:limit2}) are actually bounds
on a set of linear combination of gLIV coefficients. Now we improve the
analysis method to a global approach, and extend the study to the whole
GWTC-1 catalog. Due to the presence of multiple events, we are privileged
to carry out a global analysis that breaks the degeneracy of various gLIV
parameters. Such a global analysis was not possible for the time being of
Ref.~\cite{Kostelecky:2016kfm} with only GW150914 detected then.

%---------------------------------------------------------------------
\begin{figure}[ht]
  \centering\includegraphics[width=8cm]{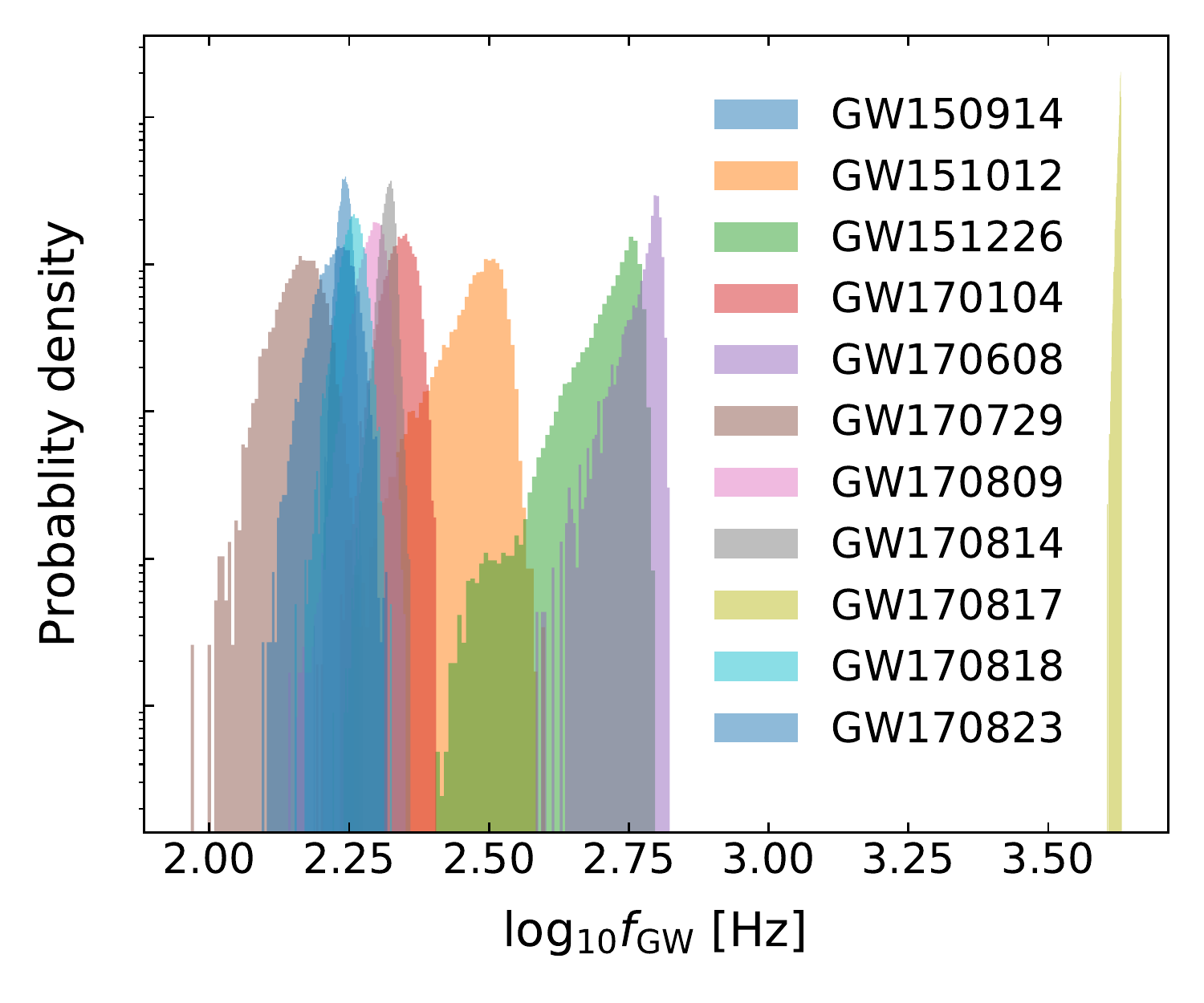}
\caption{Probability density for $f_{\rm GW}$, generated using the posterior 
samples provided by the LIGO/Virgo Collaboration~\cite{LIGOScientific:2018mvr,
GWTC1:catalog, GWTC1:PE} and the fit in
Eq.~(\ref{eq:fGW:peak})~\cite{Bohe:2016gbl}. \label{fig:fGW}}
\end{figure}
%---------------------------------------------------------------------

In order to construct the gLIV tests with Eq.~(\ref{eq:Delta:ta}) in
practice, we make the following considerations.
%--
\begin{enumerate}[(i)]
  \item For the same reason as that in the photon
  sector~\cite{Kostelecky:2002hh, Kostelecky:2009zp, Shao:2011uc}, in these
  propagation tests vacuum birefringent phenomena can constrain gLIV
  parameters more tightly than the dispersive ones. Because
  $\zeta^0_{(d)}$, so as $\tilde \zeta^0_{(d)}$, introduces {\it
  polarization-independent} time delays, they are comparably loosely bound.
  Nevertheless, if GW companion particles (photons or neutrinos) are
  detected, the dispersion can be fairly constrained; for example, see the
  bound on the SME $\bar s_{00}^{(4)}$ parameter via the simultaneous
  observation of GW170817 and GRB\,170817A~\cite{Monitor:2017mdv}. To bound
  the scope of this paper with fair workload, these
  polarization-independent, {\it dispersive-only} delays, encoded in
  $\zeta^0_{(d)}$ and $\tilde \zeta^0_{(d)}$, are omitted in the following
  analysis. They can be incorporated in future work.
  \item For the GW frequency at the amplitude peak, we use the dedicated
  fit for the $(2,2)$ mode in the Appendix A.3 of Ref.~\cite{Bohe:2016gbl}.
  The fit was obtained by combining catalogs of numerical
  relativity~\cite{Boyle:2019kee} and test-particle Teukolsky-code
  waveforms. It includes the contribution from the mass ratio of the binary
  and the orbit-aligned spins of the binary components. The fit is adopted
  in the so-called SEOBNRv4 waveform family~\cite{Bohe:2016gbl}, and is
  being extensively applied in the LIGO/Virgo daily data analysis. An
  explicit expression for the global fit is given in
  Appendix~\ref{app:peak:frequency} and several examples for different spin
  combinations are given in Fig.~\ref{fig:peak:frequency} therein. 
  Different waveform families give very close, practically
  indistinguishable results~\cite{Abbott:2016wiq}.
  Using
  the posterior samples provided by the LIGO/Virgo
  Collaboration~\cite{LIGOScientific:2018mvr, GWTC1:catalog, GWTC1:PE}, we
  plot the GW peak frequencies for the 11 GW events in
  Fig.~\ref{fig:fGW}. Notice that, while the fit was obtained from BBH
  simulations without matter effects, its prediction to the BNS event
  GW170817, whose nuclear matter effects enter the waveform at the fifth
  post-Newtonian order~\cite{Hinderer:2008dm, Abbott:2018wiz}, should be
  indicative for the analysis in this work. Because the merger part for BNS
  was not actually observed in GW
  detectors~\cite{TheLIGOScientific:2017qsa, Abbott:2018wiz}, in our tests
  we conservatively use a rather arbitrary value $f_{\rm GW} = 800$\,Hz for
  GW170817. It roughly corresponds to the cutoff sensitivity at
  high-frequency end for LIGO/Virgo detectors at the time of GW170817.
  Larger values of $f_{\rm GW}$ would lead to tighter constraints.
  \item For the time delay between two circularly polarized GW modes, we
  use a simple {\it order-of-magnitude} estimation, $\left| \omega_{\rm
  GW}\Delta t \right| \leq 2\pi / \rho$, where $\Delta t$ is the time
  difference between two circular modes, and $\rho$ is the network
  signal-to-noise ratio (SNR) of the observed GW events (see
  Table~\ref{tab:gwtc1} in Appendix~\ref{app:gwtc}). We expect this
  estimation to work fairly well at the current stage in {\it constraining}
  the gLIV phenomena, instead of {\it discovering} the gLIV phenomena. One
  direction to improve the investigation would be using the
  matched-filtering technique with modified/deformed gravitational
  waveforms~\cite{Mewes:2019dhj}. For such an improved study, our results
  from individual events can be rescaled accordingly. Study along this line
  can be used to verify our assumption. For now we leave a dedicated
  analysis to future work.
  %--
  \item Limited by the number of available GW events we will focus on (i)
  the mass dimension $d=5$ gLIV coefficients $k^{(5)}_{(V)jm}$, and (ii)
  the mass dimension $d=6$ gLIV coefficients $k^{(6)}_{(E)jm}$ and
  $k^{(6)}_{(B)jm}$. As mentioned before, mathematically it is required
  that $|s | \leq j \leq d-2$; therefore, when $d=5$, $j=0,1,2,3$ for
  $k^{(5)}_{(V)jm}$, and when $d=6$, $j$ only takes the value $4$ for
  $k^{(6)}_{(E)jm}$ and $k^{(6)}_{(B)jm}$. In general, the components of
  $\left\{ k^{(d)}_{(V)jm}, k^{(d)}_{(E)jm}, k^{(d)}_{(B)jm} \right\} $ are
  complex numbers, satisfying $\left(k_{j m}^{(d)}\right)^*=(-1)^{m}
  k_{j(-m)}^{(d)}$~\cite{Mewes:2019dhj}. Thus, we will deal with in total,
  (i) $(d-1)^2 = 16$ independent components for $k^{(5)}_{(V)jm}$, and (ii)
  $(d-1)^2-16=9$ independent components for $k^{(6)}_{(E)jm}$ and
  $k^{(6)}_{(B)jm}$ each.
\end{enumerate}
%--

In a short summary, on one hand we have 11 two-sided constraints from 11 GW
events in the GWTC-1, and on the other hand, we need to constrain $16$
independent components when $d=5$, and $9+9=18$ independent components when
$d=6$. Therefore, it is an {\it over-constraining} system. The reason for
the GW propagation constraints being ``two sided'' is that, we do not
expect either circular mode travels faster than the other one; otherwise,
the deformation in the waveforms would be quite
obvious~\cite{Mewes:2019dhj}.

%---------------------------------------------------------------------
\def\arraystretch{1.8}
\begin{table}[t]
\caption{The {\it maximal-reach} 1-$\sigma$ limits on the magnitude of
$k^{(5)}_{(V)jm}$ by assuming that the other gLIV coefficients are zero.
The tightest constraints from an {\it individual} GW event are listed
alongside with the event names. The {\it combined} constraints are obtained
via $\sigma^{\rm (combined)} \equiv 1 / \sqrt{ \sum_{i}^{N_{\rm GW}}
1/\sigma_i^2} $. All limits are given in unit of $10^{-16}\,{\rm m}$.
\label{tab:d5:individual}} 
\centering
\begin{tabular}{p{0.5cm}p{0.6cm}p{1.9cm}p{1.5cm}p{2cm}p{1.4cm}}
\hline\hline
$j$ & $m$ & Component & Individual & GW & Combined \\
\hline
0 & 0 & $\left| k^{(5)}_{(V)00} \right|$ & $5.3$ & GW170608 & 3.3 \\
1 & 0 & $\left| k^{(5)}_{(V)10} \right|$ & $3.5$ & GW170608 & 2.7 \\
1 & 1 & $\left| {\rm Re}\, k^{(5)}_{(V)11} \right|$ & $8.9$ & GW151226 & 4.7 \\
 &  & $\left| {\rm Im}\, k^{(5)}_{(V)11} \right|$  & $8.9$ & GW151226 & 4.7 \\
2 & 0 & $\left| k^{(5)}_{(V)20} \right|$ & $3.8$ & GW170608 & 3.1 \\
2 & 1 & $\left| {\rm Re}\, k^{(5)}_{(V)21} \right|$ & $4.9$ & GW170608 & 3.4 \\
 &  & $\left| {\rm Im}\, k^{(5)}_{(V)21} \right|$ & $4.9$ & GW170608 & 3.4 \\
2 & 2 & $\left| {\rm Re}\, k^{(5)}_{(V)22} \right|$ & $8.6$ & GW170817 & 5.8 \\
 &  & $\left| {\rm Im}\, k^{(5)}_{(V)22} \right|$  & $8.6$ & GW170817 & 5.8 \\
3 & 0 & $\left| k^{(5)}_{(V)30} \right|$  & $6.8$ & GW151226 & 3.9 \\
3 & 1 & $\left| {\rm Re}\, k^{(5)}_{(V)31} \right|$ & $3.7$ & GW170608 & 3.1 \\
 &  & $\left| {\rm Im}\, k^{(5)}_{(V)31} \right|$  & $3.7$ & GW170608 & 3.1 \\
3 & 2 & $\left| {\rm Re}\, k^{(5)}_{(V)32} \right|$  & $7.8$ & GW170608 & 4.4 \\
 &  & $\left| {\rm Im}\, k^{(5)}_{(V)32} \right|$  & $7.8$ & GW170608 & 4.4 \\
3 & 3 & $\left| {\rm Re}\, k^{(5)}_{(V)33} \right|$ & $8.3$ & GW170817 & 6.5 \\
 &  & $\left| {\rm Im}\, k^{(5)}_{(V)33} \right|$ & $8.3$ & GW170817 & 6.5 \\
\hline
\end{tabular}
\end{table}
%---------------------------------------------------------------------

%---------------------------------------------------------------------
\def\arraystretch{1.8}
\begin{table}[t]
\caption{Same as Table~\ref{tab:d5:individual}, for $k^{(6)}_{(E)jm}$ and
$k^{(6)}_{(B)jm}$. All limits are given in unit of $10^{-11}\,{\rm m}^2$.
\label{tab:d6:individual} } 
\centering
\begin{tabular}{p{0.5cm}p{0.6cm}p{1.9cm}p{1.5cm}p{2cm}p{1.4cm}}
  \hline\hline
  $j$ & $m$ & Component & Individual & GW & Combined \\
\hline
4 & 0 & $\left| k^{(6)}_{(E,B)40} \right|$ & $4.9$ & GW170817 & 4.8 \\
4 & 1 & $\left| {\rm Re}\, k^{(6)}_{(E,B)41} \right|$ & $5.0$ & GW170817 & 4.6 \\
 &  & $\left| {\rm Im}\, k^{(6)}_{(E,B)41} \right|$ & $5.0$ & GW170817 & 4.6 \\
4 & 2 & $\left| {\rm Re}\, k^{(6)}_{(E,B)42} \right|$ & $5.7$ & GW170817 & 4.1 \\
 &  & $\left| {\rm Im}\, k^{(6)}_{(E,B)42} \right|$ & $5.7$ & GW170817 & 4.1 \\
4 & 3 & $\left| {\rm Re}\, k^{(6)}_{(E,B)43} \right|$ & $3.9$ & GW170608 & 3.0 \\
 &  & $\left| {\rm Im}\, k^{(6)}_{(E,B)43} \right|$ & $3.9$ & GW170608 & 3.0 \\
 4 & 4 & $\left| {\rm Re}\, k^{(6)}_{(E,B)44} \right|$ & $2.7$ & GW170608 & 2.4 \\
 &  & $\left| {\rm Im}\, k^{(6)}_{(E,B)44} \right|$ & $2.7$ & GW170608 & 2.4 \\
\hline
\end{tabular}
\end{table}
%---------------------------------------------------------------------

%---------------------------------------------------------------------
\section{Numerical resutls}
\label{sec:res}
%---------------------------------------------------------------------

With all the practical considerations in Secs.~\ref{sec:dispersion} and
\ref{sec:bire} taken into account, we present our final numerical
constraints on the gLIV coefficients in this section. As mentioned above,
it is an over-constraining system. Therefore, we can consequently bound
{\it globally} all independent coefficients at a specific mass dimension
$d$ using the whole GW catalog. However, we will first consider the {\it
maximal-reach} scenario~\cite{Tasson:2019kuw, Kostelecky:2008ts} where only
one independent component is assumed to be nonzero. It eases comparison
with results in Ref.~\cite{Kostelecky:2016kfm}, and provides us some
insight on the figure of merit for different GW events in testing the
anisotropic birefringence.

Because the GW parameters are generally highly correlated, we use the
posterior samples provided by the LIGO/Virgo
Collaboration~\cite{GWTC1:catalog, GWTC1:PE} which include all correlations
among parameters and are publicly available. For the ten BBHs, we use the
posterior samples tagged with {\sc Overall\_posterior} that are combined
results derived from the effective-one-body and phenomenological waveform
families~\cite{LIGOScientific:2018mvr}, while for the BNS GW170817, we use
the samples tagged with {\sc IMRPhenomPv2NRT\_lowSpin\_posterior} that are
derived from an effectively-precessing phenomenological waveform family
with the tidal deformability effects incorporated~\cite{Abbott:2018wiz}.
Other choices do not change our limits in a significant way. The GW
parameters we use include, (i) the sky position represented by the right
ascension, $\alpha$, and the declination, $\delta$ (see
Fig.~\ref{fig:skymap} in Appendix~\ref{app:gwtc}); (ii) the intrinsic GW
parameters including the component masses, $m_1$ and $m_2$, and the
(dimensionless) orbital-aligned component spins, $\chi_1$ and $\chi_2$; and
(iii) the luminosity distance $d_L$ from where the redshift, and then the
$D_\alpha^{(d)}$ in Eq.~(\ref{eq:Dalpha}), are derived with the standard
$\Lambda$CDM model~\cite{Aghanim:2018eyx}.

In the {\it maximal-reach} scenario, we assume that only one gLIV parameter
is nonzero at a time~\cite{Tasson:2019kuw}. For each GW event, we randomly
draw samples from the posteriors. We calculate the individual limit for
each gLIV parameter for each sample point. The results are stored for
statistical inference afterwards. After accumulating enough samples, the
1-$\sigma$ limits are obtained from their corresponding distributions. In
the calculation, all statistical uncertainties are taken into account
properly.

In Tables~\ref{tab:d5:individual} and \ref{tab:d6:individual}, respectively
for mass dimensions $d=5$ and $d=6$, we list the {\it maximal-reach} limits
for each gLIV component. We have put the tightest limit from an individual
GW event, along with the event name in the fourth and fifth columns. We can
see that, for mass dimension-5 operators, the individually tightest limits
come from GW151226, GW170608, and GW170817, while for mass dimension-6
operators, the individually tightest limits come from GW170608 and
GW170817. These events all have low component masses. 

Usually, low-mass events are detected closer to the Earth, because their GW
strain amplitudes are smaller than high-mass ones if they were put at a
same distance~\cite{Maggiore:1900zz, Aasi:2013wya}. The finite sensitivity
of GW detectors can only pick up those low-mass events relatively
nearby~\cite{Aasi:2013wya}. For the three individually best events here
(GW151226, GW170608, and GW170817), they all have the cosmological redshift
$z \lesssim 0.1$ (see Table~\ref{tab:gwtc1}). Although the GW propagation
tests benefit from large distances [see Eq.~(\ref{eq:Dalpha})], which
usually correspond to high-mass GW events (see Table~\ref{tab:gwtc1}), our
anisotropic birefringent tests also depend on the GW frequency with a
powerlaw index $d-4$ [see Eq.~(\ref{eq:Delta:ta})], which is positive for
$d=5$ and $6$. High-mass GW events have a lower GW frequency [see
Eq.~(\ref{eq:fGW:peak})]~\cite{Maggiore:1900zz}, which decreases their
ability to constrain the gLIV parameters. Our results clearly show that,
already at mass dimension 5, low-mass events are preferred to the tests of
the gLIV parameters in the SME. At mass dimension 6, the benefit from
smaller masses is even more prominent. The dependence of the bounds
on other parameters, e.g. the precision of sky localization, deserves
further study.

For the {\it maximal-reach} scenario, we also list the combined limits from
multiple GWs by assuming each event mutually independently. Therefore, one
gets a combined limit $\sigma^{\rm (combined)} \simeq 1/
\sqrt{\sum_i^{N_{\rm GW}} 1/\sigma_i^2 }$, where $\sigma_i$ is the limit
from an individual GW event $i$. As we can see, the combined limit improves
the limit from the individually best event by a factor smaller than two.
Therefore, in the {\it maximal-reach} approach, the limit is dominated by
the individually best GW event. Worth to mention that, the limits in
Tables~\ref{tab:d5:individual} and \ref{tab:d6:individual} improve over the
first set of limits from GW150914~\cite{Kostelecky:2016kfm} by a factor
$\sim 10^2$--$10^4$ for the mass dimension $d=5$, and by a factor $\sim
10^3$--$10^5$ for the mass dimension $d=6$.

%---------------------------------------------------------------------
\begin{figure*}[htp]
  \centering\includegraphics[width=18cm]{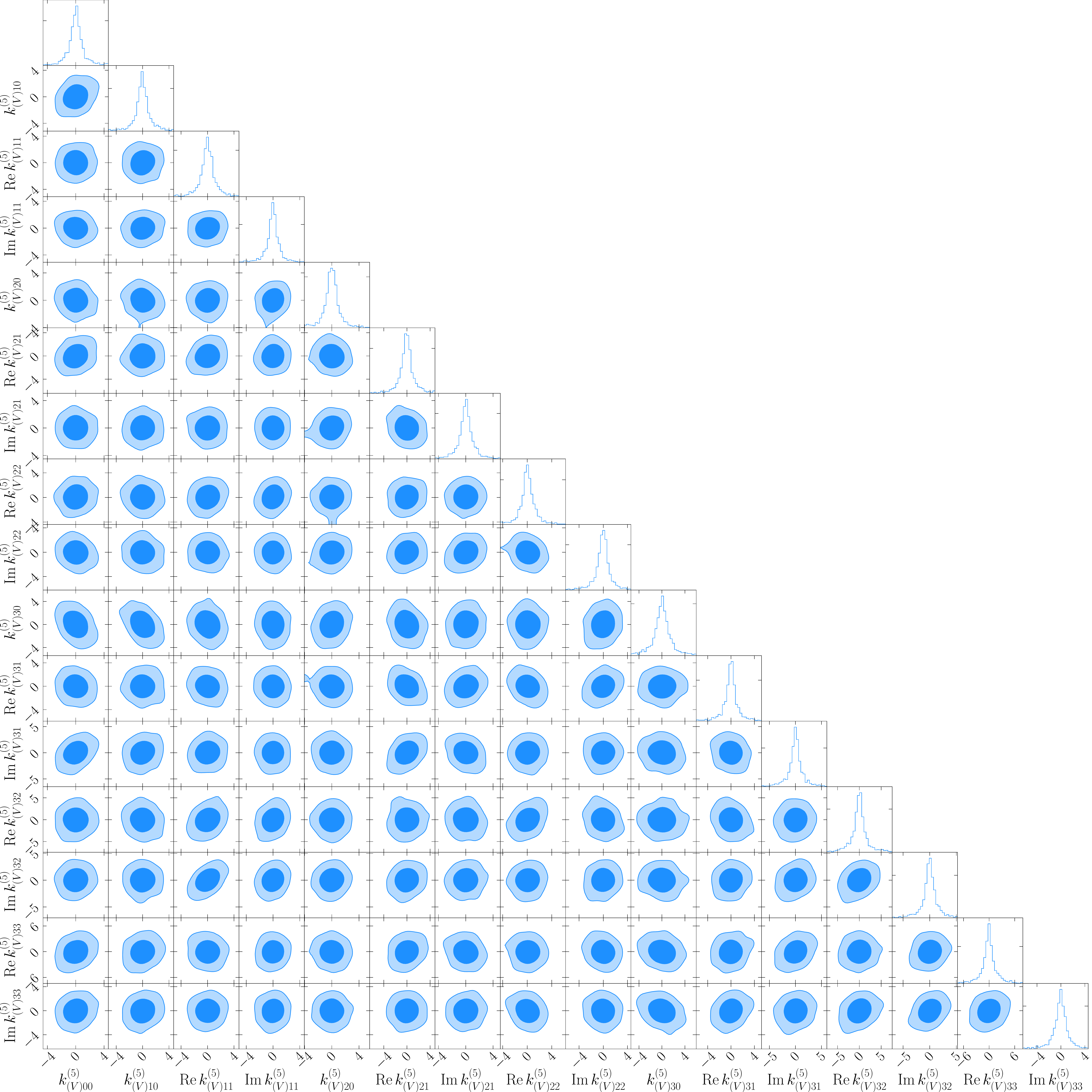}
\caption{Global constraints on 16 independent gLIV components with mass dimension
$d=5$ from 11 GW events in the GWTC-1. Contours show the 68\% and 90\%
confidence levels. The unit for the limits is $10^{-15}\,{\rm m}$ in this
figure. \label{fig:global:d5}}
\end{figure*}
%---------------------------------------------------------------------
  
%---------------------------------------------------------------------
\begin{figure*}[htp]
\centering\includegraphics[width=18cm]{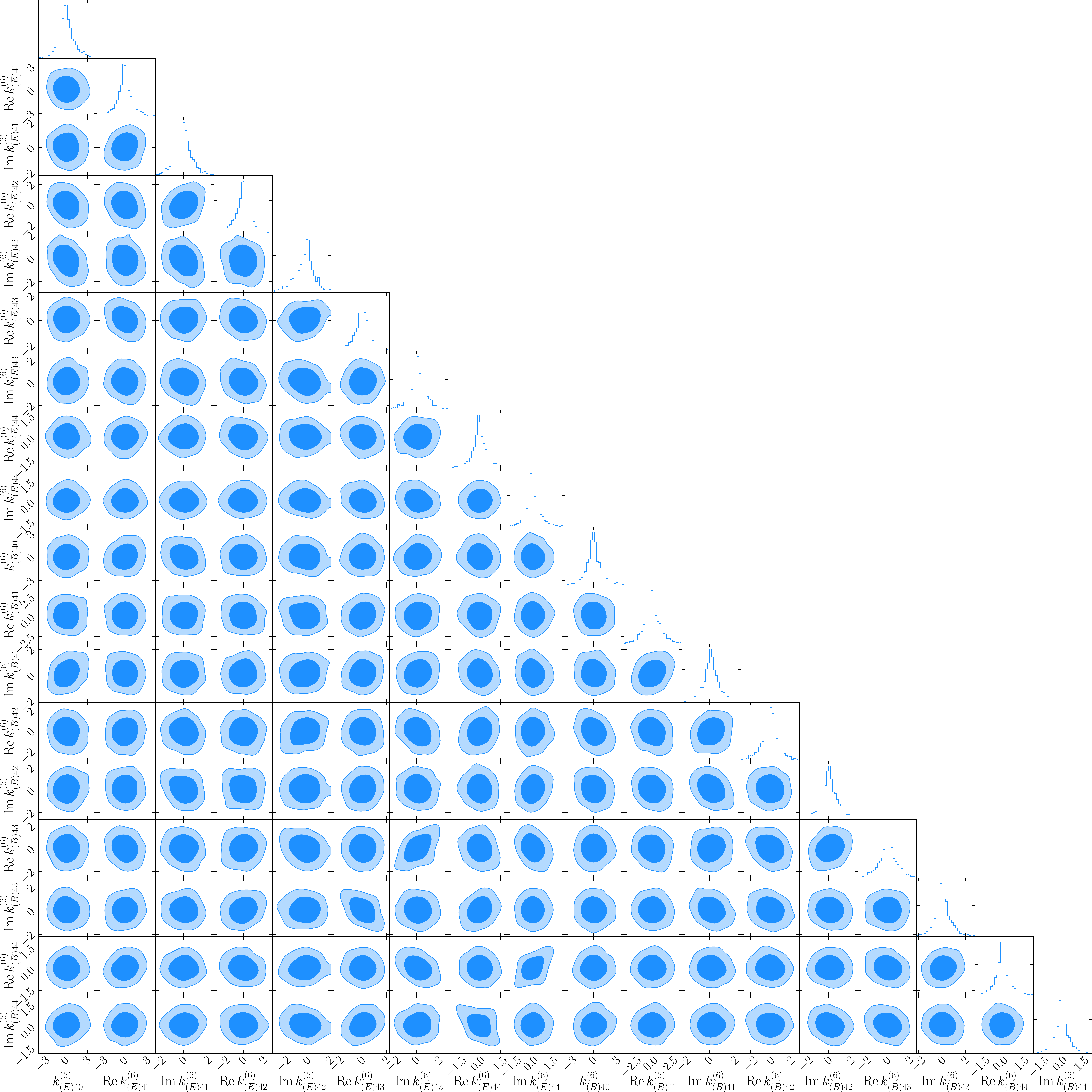}
\caption{Same as Fig.~\ref{fig:global:d5}, for the 18 independent gLIV components
with mass dimension $d=6$. The unit for the limits is $10^{-10}\,{\rm m}^2$
in this figure. \label{fig:global:d6}}
\end{figure*}
%---------------------------------------------------------------------

The {\it maximal-reach} limits should be considered as quite optimistic,
and in reality they should be bounds on some linear combinations of the
underlying set of gLIV coefficients~\cite{Kostelecky:2016kfm}. Therefore,
it becomes very intriguing to break the degeneracy among these parameters
and check the real power of GW events in probing the gLIV when a number of
events are available~\cite{Kostelecky:2008ts}. In the following, we use the
simple facts, that (i) different GW events come from different
directions~\cite{LIGOScientific:2018mvr} and (ii) symmetry breaking in the
SME is observer Lorentz-invariant~\cite{Bluhm:2004ep}, to {\it globally}
constrain these gLIV parameters. This is possible, because the coefficients
of the linear combinations are direction-dependent, depending on the
direction $\hat{\bm{n}}$ via the spin-weighted spherical harmonics
functions; see Eqs.~(\ref{eq:kI}--\ref{eq:kV}) and
Appendix~\ref{app:spin:harmonics}.

%---------------------------------------------------------------------
\def\arraystretch{1.5}
\begin{table}[t]
\caption{{\it Global} constraints on $k^{(5)}_{(V)jm}$ at the 68\% confidence 
level. Notice that the unit differs from that in Table~\ref{tab:d5:individual}.
\label{tab:d5:global}}
\centering
\begin{tabular}{p{1cm}p{1cm}p{2cm}p{3cm}}
\hline\hline
$j$ & $m$ & Component & Constraint [$10^{-15}\,{\rm m}$] \\
\hline
0 & 0 & $k^{(5)}_{(V)00}$ & $(-0.94, 0.96)$  \\
1 & 0 & $k^{(5)}_{(V)10}$ & $(-1.01, 1.03)$ \\
1 & 1 & ${\rm Re}\, k^{(5)}_{(V)11}$ & $(-0.99, 1.06)$ \\
 &  & ${\rm Im}\, k^{(5)}_{(V)11}$  & $(-0.89, 0.84)$ \\
2 & 0 & $k^{(5)}_{(V)20}$ & $(-1.00, 0.94)$ \\
2 & 1 & ${\rm Re}\, k^{(5)}_{(V)21}$ & $(-1.12, 1.08)$ \\
 &  & ${\rm Im}\, k^{(5)}_{(V)21}$ & $(-0.96, 1.00)$ \\
2 & 2 & ${\rm Re}\, k^{(5)}_{(V)22}$ & $(-0.97, 1.19)$ \\
 &  & ${\rm Im}\, k^{(5)}_{(V)22}$  & $(-1.13, 0.93)$ \\
3 & 0 & $k^{(5)}_{(V)30}$  & $(-1.38, 1.43)$ \\
3 & 1 & ${\rm Re}\, k^{(5)}_{(V)31}$ & $(-1.13, 0.98)$ \\
 &  & ${\rm Im}\, k^{(5)}_{(V)31}$  & $(-1.26, 1.25)$ \\
3 & 2 & ${\rm Re}\, k^{(5)}_{(V)32}$  & $(-1.54, 1.46)$ \\
 &  & ${\rm Im}\, k^{(5)}_{(V)32}$  & $(-1.22, 1.14)$ \\
3 & 3 & ${\rm Re}\, k^{(5)}_{(V)33}$ & $(-1.45, 1.50)$ \\
 &  & ${\rm Im}\, k^{(5)}_{(V)33}$ & $(-1.20, 0.98)$ \\
\hline
\end{tabular}
\end{table}
%---------------------------------------------------------------------

%---------------------------------------------------------------------
\def\arraystretch{1.5}
\begin{table}[t]
\caption{Same as Table~\ref{tab:d5:global}, for $k^{(6)}_{(E)jm}$ and
$k^{(6)}_{(B)jm}$. 
\label{tab:d6:global} } 
\centering
\begin{tabular}{p{1cm}p{1cm}p{2cm}p{3cm}}
\hline\hline
$j$ & $m$ & Component & Constraint [$10^{-10}\,{\rm m}^2$] \\
\hline
4 & 0 & $k^{(6)}_{(E)40}$ & $(-0.76, 1.32)$ \\
4 & 1 & ${\rm Re}\, k^{(6)}_{(E)41}$ & $(-0.81, 1.16)$ \\
 &  & ${\rm Im}\, k^{(6)}_{(E)41}$ & $(-0.68, 0.72)$ \\
4 & 2 & ${\rm Re}\, k^{(6)}_{(E)42}$ & $(-0.85, 0.81)$ \\
 &  & ${\rm Im}\, k^{(6)}_{(E)42}$ & $(-1.11, 0.55)$ \\
4 & 3 & ${\rm Re}\, k^{(6)}_{(E)43}$ & $(-0.57, 0.69)$ \\
 &  & ${\rm Im}\, k^{(6)}_{(E)43}$ & $(-0.58, 0.88)$ \\
 4 & 4 & ${\rm Re}\, k^{(6)}_{(E)44}$ & $(-0.39, 0.58)$ \\
 &  & ${\rm Im}\, k^{(6)}_{(E)44}$ & $(-0.32, 0.67)$ \\
 \hline
 4 & 0 & $k^{(6)}_{(B)40}$ & $(-0.87, 1.02)$ \\
 4 & 1 & ${\rm Re}\, k^{(6)}_{(B)41}$ & $(-0.82, 1.21)$ \\
  &  & ${\rm Im}\, k^{(6)}_{(B)41}$ & $(-0.47, 0.80)$ \\
 4 & 2 & ${\rm Re}\, k^{(6)}_{(B)42}$ & $(-0.93, 0.74)$ \\
  &  & ${\rm Im}\, k^{(6)}_{(B)42}$ & $(-0.63, 0.87)$ \\
 4 & 3 & ${\rm Re}\, k^{(6)}_{(B)43}$ & $(-0.70, 0.76)$ \\
  &  & ${\rm Im}\, k^{(6)}_{(B)43}$ & $(-0.55, 0.77)$ \\
  4 & 4 & ${\rm Re}\, k^{(6)}_{(B)44}$ & $(-0.39, 0.60)$ \\
  &  & ${\rm Im}\, k^{(6)}_{(B)44}$ & $(-0.34, 0.64)$ \\
\hline
\end{tabular}
\end{table}
%---------------------------------------------------------------------

We have two sets of {\it global} analysis, for mass dimension $5$ and mass
dimension $6$. In each {\it global} analysis, we assume that {\it all}
anisotropic-birefringent gLIV coefficients can be nonzero at that specific
$d$. Therefore, we have in total 16 independent coefficients for $d=5$, and
18 independent coefficients for $d=6$. Fortunately, the 11 events in GWTC-1
provide us in total 22 useful constraints.

The global analysis for $d=5$ is relatively easier, because (i) the square
root in Eq.~(\ref{eq:Delta:ta}) is plainly calculated when
$\tilde{\zeta}_{(d=5)}^{1} = \tilde{\zeta}_{(d=5)}^{2} = 0$, and (ii) only
the $s=0$ spherical harmonics are involved. Similarly to the maximal-reach
case, we randomly draw posterior samples, but now simultaneously for 11
events. Their time delays are drawn from zero-mean Gaussian distributions
with their standard variances determined in Sec.~\ref{sec:bire}. Then, for
each random draw, we construct the global likelihood as a function of the
16 gLIV coefficients. The likelihood is maximized by the routines in the
{\sc Scipy.Optimize} package~\cite{Virtanen:2019joe}. Thus, values of 16
coefficients are calculated in each random draw. The resulted values for
the 16 gLIV coefficients are recorded for later inference. After we
accumulate enough draws, we extract the constraints on the 16 gLIV
coefficients from these recorded distributions. 

The 1-dimensional marginalized constraints are listed in
Table~\ref{tab:d5:global}, and the contour plots for these parameters are
given in Fig.~\ref{fig:global:d5}. It is interesting to note that, (i) the
{\it global} constraints are only about a factor of $10$ weaker than the
{\it maximal-reach} limits, and (ii) these 16 gLIV parameters are hardly
correlated. The small correlation between parameters is resulted from the
use of multiple events. It shows the advantage of constructing an {\it
over-constraining} system with multiple events, analogous to the cases of
using millisecond pulsars under a similar context~\cite{Shao:2014oha,
Shao:2018vul, Shao:2019cyt}.

The global analysis for $d=6$ is somehow complicated. Now we have the
spin-weighted spherical harmonics with $s=\pm 4$. To reduce the
computational cost, we use $_{s} Y_{j m}^{*}={(-1)^{s+m}} \, {_{-s}
Y_{j(-m)}}$ to calculate ${_{-4} Y_{jm}}$ from ${_{+4} Y_{jm}}$. More
importantly, in the most general case for $d=6$, we have
$\tilde{\zeta}_{(d=6)}^{1} \neq 0$ and $\tilde{\zeta}_{(d=6)}^{2} \neq 0$.
As a consequence, the square root in Eq.~(\ref{eq:Delta:ta}) is nontrivial.
Therefore, the calculation with these highly nonlinear features takes a
much longer computer time, namely, the {\sc Scipy.Optimize} routines now
need much more significant computational time to iterate, in order to
locate the maximum of the 18-dimensional likelihood functions.
Nevertheless, the principles are similar to the $d=5$ case. The results for
$d=6$ are give in Table~\ref{tab:d6:global} and Fig.~\ref{fig:global:d6}.
Similar conclusions that were made for $d=5$ operators in the last
paragraph can be made for $d=6$ operators as well.

%---------------------------------------------------------------------
\section{Discussion}
\label{sec:diss}
%---------------------------------------------------------------------

Gravitational Lorentz invariance violation (gLIV) is a central topic in the
program of using the gravitational waves (GWs) to probe the most
fundamental principles in modern physics~\cite{Abbott:2017vtc,
LIGOScientific:2019fpa, Kostelecky:2016kfm, Nishizawa:2017nef,
Yunes:2016jcc, Berti:2018cxi, Berti:2018vdi}. If the Lorentz symmetry is
broken in the gravity sector, there might be a preferred
frame~\cite{Will:2014kxa,Will:2018bme} where the breaking is isotropic, but
a more generic breaking does not need a preferred
frame~\cite{Kostelecky:2003fs, Bailey:2006fd}. Isotropic gLIV, being a
specific class of gLIV, was studied in details~\cite{Mirshekari:2011yq,
Abbott:2017vtc, LIGOScientific:2019fpa, Wang:2020pgu, Nishizawa:2017nef}.
Most of previous work focused on non-birefringent phenomena in searches of
the gLIV. Extensions to these studies are needed. On one hand, because the
commutator of two boost generators is a generator for rotation, anisotropy
is inevitable in a complete search for the gLIV. On the other hand, in
effective field theories, the most general gLIV has birefringent behaviors
for the two circularly polarized GW eigen-modes in the nonminimal gravity.
In this work, we take a leap to systematically study anisotropic
birefringent phenomena related to the GW propagation with the GW transient
catalog GWTC-1~\cite{LIGOScientific:2018mvr, GWTC1:catalog, GWTC1:PE}.

One of the best theoretical frameworks, in carrying out these anisotropic
birefringent gLIV tests, is the standard-model extension
(SME)~\cite{Kostelecky:2003fs, Kostelecky:2016kfm, Mewes:2019dhj}. We
follow the spirit of effective field theories, and assume that the gLIV
happens at a specific mass dimension $d$. The lowest mass dimensions for
vacuum birefringence are $d=5$ and $d=6$. When $d=5$, there are in total 16
independent gLIV coefficients in $k_{(V) j m}^{(5)}$, while $d=6$, there
are 18 independent gLIV coefficients in $k_{(E) j m}^{(6)}$ and $k_{(B) j
m}^{(6)}$. In our {\it global} tests, we simultaneously include {\it all}
gLIV operators that lead to birefringence at that particular mass
dimension. We use the posterior samples for the events in GWTC-1, provided
by the LIGO/Virgo Collaboration, to coherently solve for gLIV parameters,
and obtain their constraints thereof. Our {\it maximal-reach} limits are
listed in Tables~\ref{tab:d5:individual} and \ref{tab:d6:individual}, and
our {\it global} limits are presented in Tables~\ref{tab:d5:global} and
\ref{tab:d6:global}, as well as in Figs.~\ref{fig:global:d5} and
\ref{fig:global:d6}. No violation of Einstein's general relativity was
found, and the constraints on 34 gLIV coefficients are improved by factors
ranging from $\sim 10^2$ to $\sim 10^5$, with respect to previous
limits~\cite{Kostelecky:2016kfm}.

The gLIV tests can be improved in multiple directions in the future. (I) A
more sophisticated matched-filtering analysis with gLIV-deformed
waveforms~\cite{Mewes:2019dhj} can be used to validate the assumptions made
in this work, though such an analysis could have cost mighty computational
time in practice for a catalog of GWs. (II) Another possibility in testing
the GW propagation can involve modified cosmological behaviors. For
example, \citet{Nishizawa:2017nef} considered a generic GW propagation
equation,
%--
\begin{align}\label{eq:cosmo}
  h_{ij}^{\prime\prime} + \left( 2+\nu \right) {\cal H} h_{ij}^\prime + \left(
  c_T^2 k^2 + a^2 \mu^2 \right) h_{ij} &= a^2 \Gamma\gamma_{ij} \,,
\end{align}
%--
which --- besides the cosmological expansion encoded in ${\cal H} \equiv
a'/a$ with $a$ being the cosmological scale factor --- includes running of
the Planck mass $M_*$ via $\nu={\cal H}^{-1} d \ln M_*^2 / dt$, the
velocity of GWs $c_T$, the mass of graviton $\mu$, and the anisotropic {\it
source} term $\Gamma\gamma_{ij}$. The philosophy of the approach
(\ref{eq:cosmo}) is different from ours which is based on a modified action
of the gravity sector [see Eq.~(\ref{eq:L})]. Nevertheless, a grander
theoretical framework that includes both a modified-gravity action and a
modified cosmology might be feasible. It lays beyond the scope of this work
however. (III) The final obvious direction to improve the tests in this
work is to involve more GW events. Actually, more and yet more
accurately-measured GW events are undoubtfully
expected~\cite{Aasi:2013wya}. With the ongoing third observing run by the
LIGO/Virgo Collaboration, more GW event candidates are already revealed for
possible electromagnetic followups~\cite{O3:gracedb}. At the time of
writing, a second BNS, GW190425, was published~\cite{Abbott:2020uma}. This
event is weaker than GW170817, but it will help in constraining the gLIV
parameters in the global analysis. With the KAGRA~\cite{Aso:2013eba} and
IndiGO GW detectors coming online in the near future, even better limits
will be placed on the gLIV. Ultimately, we hope some positive clues to the
long-sought quantum gravity theory might be uncovered via GWs.

%---------------------------------------------------------------------
\def\arraystretch{1.5}
\begin{table*}[t]
\caption{Source parameters for 11 confident detections in the GWTC-1, from the first and second LIGO/Virgo observing runs~\cite{LIGOScientific:2018mvr}. Uncertainties are from the symmetric 90\% credible intervals. For the network SNR, we list the average from three detection pipelines (PyCBC, GstLAL, cWB; see Ref.~\cite{LIGOScientific:2018mvr} for details). \label{tab:gwtc1}} 
\centering
\begin{tabular}{p{1.6cm}p{1cm}p{1.5cm}p{1.5cm}p{1.5cm}p{1.8cm}p{3.2cm}p{2cm}}%{llllllll}
\hline\hline
& Type & $m_1 ~ [M_\odot]$ & $m_2 ~ [M_\odot]$ & $d_L ~ [{\rm Mpc}]$ & Redshift $z$ & Localization $\Delta\Omega~[{\rm deg}^2]$ & Network SNR \\
\hline
GW150914 & BBH & $35.6^{+4.8}_{-3.0}$ & $30.6^{+3.0}_{-4.4}$ & $430^{+150}_{-170}$ & $0.09^{+0.03}_{-0.03}$ & 180 & 24.4 \\
GW151012 & BBH & $23.3^{+14.0}_{-5.5}$ & $13.6^{+4.1}_{-4.8}$ & $1060^{+540}_{-480}$ & $0.21^{+0.09}_{-0.09}$ & $1555$ & 9.8 \\
GW151226 & BBH & $13.7^{+8.8}_{-3.2}$ & $7.7^{+2.2}_{-2.6}$ & $440^{+180}_{-190}$ & $0.09^{+0.04}_{-0.04}$ & 1033 & 12.7 \\
GW170104 & BBH & $31.0^{+7.2}_{-5.6}$ & $20.1^{+4.9}_{-4.5}$ & $960^{+430}_{-410}$ & $0.19^{+0.07}_{-0.08}$ & 924 & 13.0 \\
GW170608 & BBH & $10.9^{+5.3}_{-1.7}$ & $7.6^{+1.3}_{-2.1}$ & $320^{+120}_{-110}$ & $0.07^{+0.02}_{-0.02}$ & 396 & 14.8 \\
GW170729 & BBH & $50.6^{+16.6}_{-10.2}$ & $34.3^{+9.1}_{-10.1}$ & $2750^{+1350}_{-1320}$ & $0.48^{+0.19}_{-0.20}$ & 1033 & 10.3 \\
GW170809 & BBH & $35.2^{+8.3}_{-6.0}$ & $23.8^{+5.2}_{-5.1}$ & $990^{+320}_{-380}$ & $0.20^{+0.05}_{-0.07}$ & 340 & 12.3 \\
GW170814 & BBH & $30.7^{+5.7}_{-3.0}$ & $25.3^{+2.9}_{-4.1}$ & $580^{+160}_{-210}$ & $0.12^{+0.03}_{-0.04}$ & 87 & 16.5 \\
GW170817 & BNS & $1.46^{+0.12}_{-0.10}$ & $1.27^{+0.09}_{-0.09}$ & $40^{+10}_{-10}$ & $0.01^{+0.00}_{-0.00}$ & 16 & 32.0 \\
GW170818 & BBH & $35.5^{+7.5}_{-4.7}$ & $26.8^{+4.3}_{-5.2}$ & $1020^{+430}_{-360}$ & $0.20^{+0.07}_{-0.07}$ & 39 & 11.3 \\
GW170823 & BBH & $39.6^{+10.0}_{-6.6}$ & $29.4^{+6.3}_{-7.1}$ & $1850^{+840}_{-840}$ & $0.34^{+0.13}_{-0.14}$ & 1651 & 11.1 \\
\hline
\end{tabular}
\end{table*}
%---------------------------------------------------------------------

%---------------------------------------------------------------------
\begin{figure*}[t]
  \centering\includegraphics[trim= 0cm 3cm 0cm 2cm, width=0.8\linewidth, clip]{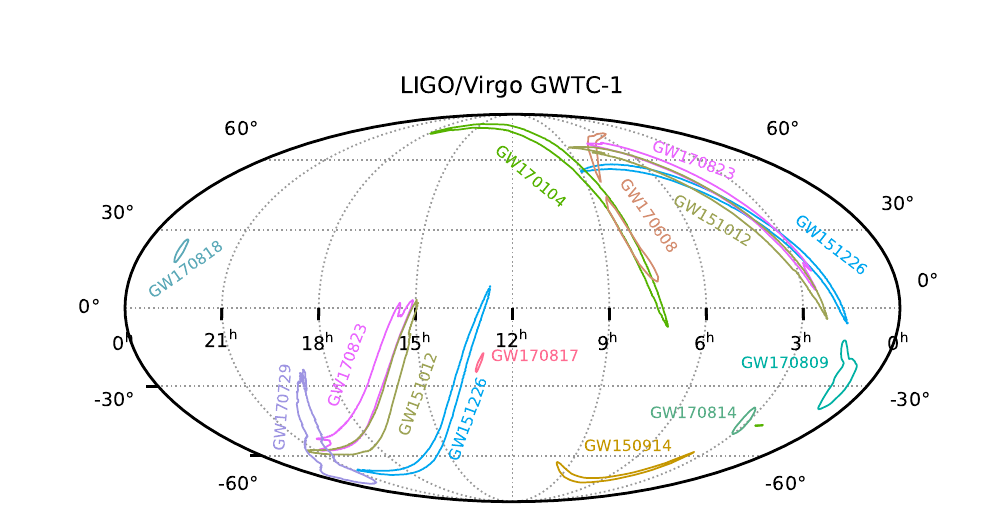}
\caption{Sky localization of GW events in the GWTC-1 catalog from
LIGO/Virgo's first and second observing runs, at the 68\% confidence
level~\cite{LIGOScientific:2018mvr}. \label{fig:skymap}}
  \end{figure*}
  %---------------------------------------------------------------------

%---------------------------------------------------------------------
\acknowledgments

We are grateful to Tjonnie Li, Alan Kosteleck\'y, and Rui Xu for helpful
discussions, and the LIGO/Virgo Collaboration for providing the posterior
samples of their parameter-estimation studies. This work was supported by the
National Natural Science Foundation of China (11975027, 11991053,
11721303), and the Young Elite Scientists Sponsorship Program by the China
Association for Science and Technology (2018QNRC001). It was partially
supported by the Strategic Priority Research Program of the Chinese Academy
of Sciences through the Grant No. XDB23010200, and the High-performance
Computing Platform of Peking University.
%---------------------------------------------------------------------

\appendix

%---------------------------------------------------------------------
\section{A brief summary of GWTC-1}
\label{app:gwtc}
%---------------------------------------------------------------------

In the Advanced LIGO/Virgo's first and second observing runs, respectively
taking place from September 12, 2015 to January 19, 2016, and from November
30, 2016 to August 25, 2017, ten confident detections of BBHs and one
confident detection of BNSs were reported~\cite{LIGOScientific:2018mvr}. In
Table~\ref{tab:gwtc1}, we list the basic parameters and their uncertainties
for these events~\cite{LIGOScientific:2018mvr}. As sky position is
important in testing anisotropic birefringence, in Fig.~\ref{fig:skymap}
we show the 68\% credible regions for sky location of GW events in the
GWTC-1~\cite{LIGOScientific:2018mvr}, in a Mollweide projection in the
equatorial coordinate. Data are taken from the associated data release for
the sky maps~\cite{GWTC1:skymaps} to the LIGO/Virgo's
GWTC-1~\cite{GWTC1:catalog}, hosted by the Gravitational Wave Open Science
Center (GWOSC)~\cite{GWSOC}. The plot has made use of the {\tt ligo.skymap}
package~\cite{package:skymap}, maintained by Leo Singer.

In all of the numerical calculations of this paper, we have used the
posterior samples provided by the LIGO/Virgo Collaboration, hosted at the GWOSC~\cite{GWTC1:PE}.

Notice that the uncertainties are rather heterogeneous for different GW
events, due to the operation of GW detectors in practice at the detection
time. In particular, the sky localization plays an essential role in
determining the anisotropic behavior of gLIV, which should be taken into
full account, as is done in this work.

%---------------------------------------------------------------------
\section{Spin-weighted spherical harmonics}
\label{app:spin:harmonics}
%---------------------------------------------------------------------

As we are familiar with the ordinary spherical harmonics, $Y_{jm} \equiv
{_0Y_{jm}} $, in describing scalars' irreducible decomposition in three
dimensions, the spin-weighted spherical harmonics, $_sY_{jm}$, are widely
used to decompose other tensors with definite orbital and spin angular
momentum~\cite{Newman:1966ub, Goldberg:1966uu}.

While mathematical and physical discussions can be found respectively in
Refs.~\cite{Newman:1966ub,Goldberg:1966uu} and in the Appendix A of
Ref.~\cite{Kostelecky:2009zp}, here in our calculation we use the explicit
expression of $_sY_{jm}$ in a brute-force manner. It
reads~\cite{Newman:1966ub,Goldberg:1966uu,Kostelecky:2009zp},
%--
\begin{align}
  _{s} Y_{j m}(\theta, \phi) =& \sqrt{ \frac{2 j+1}{4 \pi} \frac{(j+m)
  !(j-m) !}{(j+s) !(j-s) !} } e^{i m \phi} \left( \sin \frac{\theta}{2}
  \right)^{2j} \nonumber \\
  & \sum_r (-1)^{j+m+s+r} C^r_{j-s} C_{j+s}^{r+s-m} \left(\cot
  \frac{\theta}{2}\right)^{2 r+s-m} \,,
\end{align}
%--
where $C_n^k$ denotes the binomial coefficients. Interested readers are
referred to the above references for topics related to raising and lowering
operators, orthogonality relation, completeness relation, physical
interpretation with respect to the angular momentum, {\it etc.}.

%---------------------------------------------------------------------
\section{Global fit to GW frequency at the merger}
\label{app:peak:frequency}
%---------------------------------------------------------------------

%---------------------------------------------------------------------
\begin{figure}[t]
  \centering\includegraphics[width=8cm]{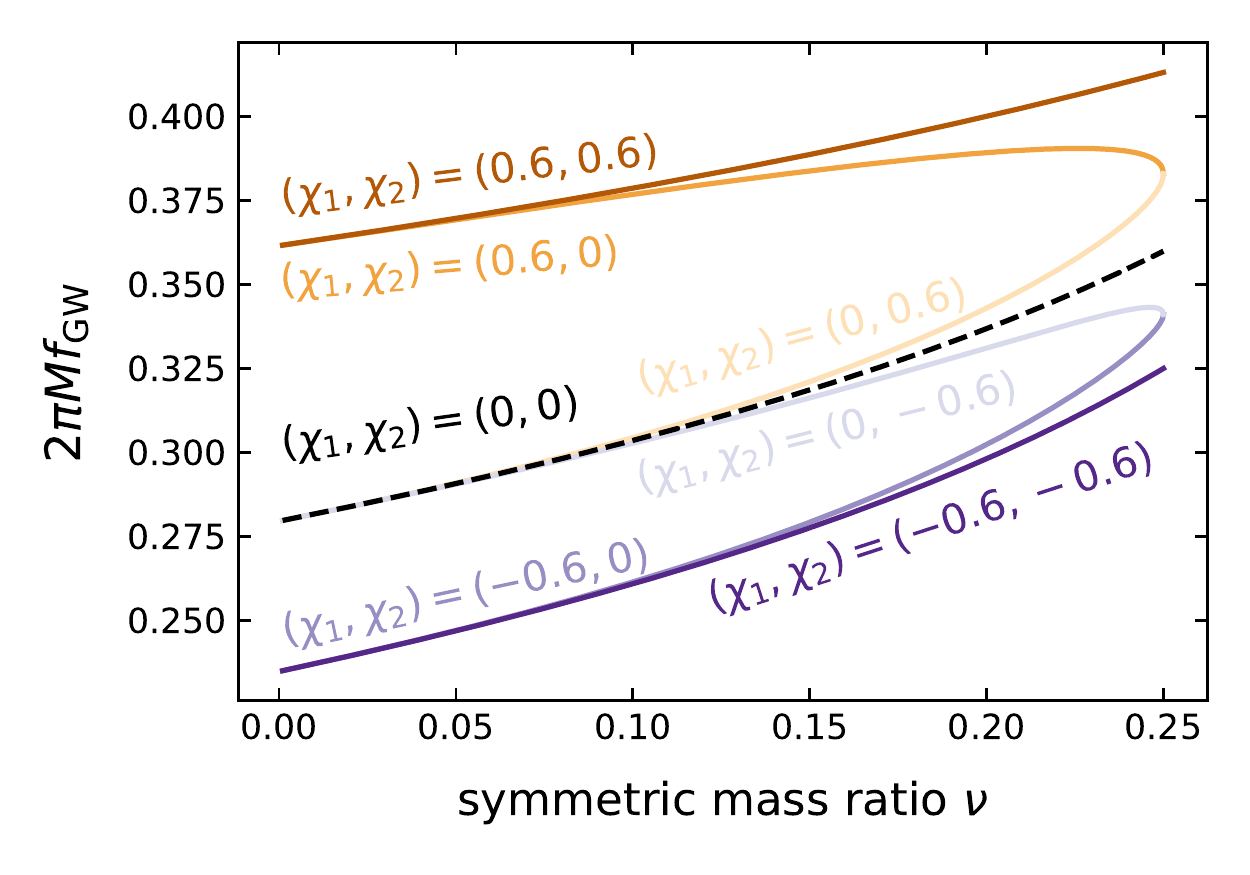}
  \caption{Examples of the GW peak frequency as a function of the symmetric mass ratio
  $\nu$, for several orbit-aligned spin combinations.
  \label{fig:peak:frequency}}
\end{figure}
%---------------------------------------------------------------------

To obtain an accurate estimation of GW frequency at the merger of a binary
system of masses $m_1$ and $m_2$ (assuming $m_1 \geq m_2$) and
(orbit-aligned) dimensionless spins $\chi_1$ and $\chi_2$, we adopt the
global fit in the Appendix A.3 of Ref.~\cite{Bohe:2016gbl}. It bases on
catalogs of waveforms from numerical relativity~\cite{Boyle:2019kee} and
test-particle Teukolsky code. Consider a binary with a symmetric mass ratio
$\nu \equiv m_1m_2/M^2$ where $M \equiv m_1 +m_2$, and an effective spin
variable,
%--
\begin{align}
  \chi \equiv \chi_{\mathrm{S}}+\frac{\chi_{\mathrm{A}}}{1-2 \nu} \delta \,,
\end{align}
%--
where $\chi_{\mathrm{S}, \mathrm{A}} \equiv\left(\chi_{1} \pm
\chi_{2}\right) / 2$, and $\delta \equiv\left(m_{1}-m_{2}\right) / M$. The
global fit for the dominant $(2,2)$ mode has the form~\cite{Bohe:2016gbl},
%--
\begin{align} \label{eq:fGW:peak}
  2 \pi Mf_{\rm GW} \left(\nu, \chi\right) = p_0 + \left(p_1 + p_2 \chi \right)
  \log \left({\cal A} - {\cal B} \chi \right) \,,
\end{align}
%--
where
%--
\begin{align}
  {\cal A} &= p_3 + 4\left( p_3 - p_4 \right) \left(\nu - \frac{1}{4}\right) \,, \\
  {\cal B} &= p_5 + 4\left( p_5 - p_6 \right) \left(\nu - \frac{1}{4}\right) \,,
\end{align}
%--
and
%--
\begin{align}
  p_0 &= 0.562679 \,, & \nonumber \\
  p_1 &= -0.087062 \,, &
  p_2 &= 0.001743 \,, \nonumber \\
  p_3 &= 10.262073 \,, &
  p_4 &= 25.850378 \,, \nonumber \\
  p_5 &= 7.629922 \,, &
  p_6 &= 25.819795 \,. \nonumber
\end{align}
%--
In Fig.~\ref{fig:peak:frequency} we show examples of the GW peak frequency
for several combinations of the orbit-aligned spins, as a function of
$\nu$. In our tests of gLIV in the main text, we have used the posteriors
of component masses and orbital-aligned component spins to infer $f_{\rm GW}$.

\clearpage

%---------------------------------------------------------------------
\bibliography{refs}

\begin{thebibliography}{88}%
\makeatletter
\providecommand \@ifxundefined [1]{%
 \@ifx{#1\undefined}
}%
\providecommand \@ifnum [1]{%
 \ifnum #1\expandafter \@firstoftwo
 \else \expandafter \@secondoftwo
 \fi
}%
\providecommand \@ifx [1]{%
 \ifx #1\expandafter \@firstoftwo
 \else \expandafter \@secondoftwo
 \fi
}%
\providecommand \natexlab [1]{#1}%
\providecommand \enquote  [1]{``#1''}%
\providecommand \bibnamefont  [1]{#1}%
\providecommand \bibfnamefont [1]{#1}%
\providecommand \citenamefont [1]{#1}%
\providecommand \href@noop [0]{\@secondoftwo}%
\providecommand \href [0]{\begingroup \@sanitize@url \@href}%
\providecommand \@href[1]{\@@startlink{#1}\@@href}%
\providecommand \@@href[1]{\endgroup#1\@@endlink}%
\providecommand \@sanitize@url [0]{\catcode `\\12\catcode `\$12\catcode
  `\&12\catcode `\#12\catcode `\^12\catcode `\_12\catcode `\%12\relax}%
\providecommand \@@startlink[1]{}%
\providecommand \@@endlink[0]{}%
\providecommand \url  [0]{\begingroup\@sanitize@url \@url }%
\providecommand \@url [1]{\endgroup\@href {#1}{\urlprefix }}%
\providecommand \urlprefix  [0]{URL }%
\providecommand \Eprint [0]{\href }%
\providecommand \doibase [0]{http://dx.doi.org/}%
\providecommand \selectlanguage [0]{\@gobble}%
\providecommand \bibinfo  [0]{\@secondoftwo}%
\providecommand \bibfield  [0]{\@secondoftwo}%
\providecommand \translation [1]{[#1]}%
\providecommand \BibitemOpen [0]{}%
\providecommand \bibitemStop [0]{}%
\providecommand \bibitemNoStop [0]{.\EOS\space}%
\providecommand \EOS [0]{\spacefactor3000\relax}%
\providecommand \BibitemShut  [1]{\csname bibitem#1\endcsname}%
\let\auto@bib@innerbib\@empty
%</preamble>
\bibitem [{\citenamefont {Jacobson}\ \emph {et~al.}(2006)\citenamefont
  {Jacobson}, \citenamefont {Liberati},\ and\ \citenamefont
  {Mattingly}}]{Jacobson:2005bg}%
  \BibitemOpen
  \bibfield  {author} {\bibinfo {author} {\bibfnamefont {T.}~\bibnamefont
  {Jacobson}}, \bibinfo {author} {\bibfnamefont {S.}~\bibnamefont {Liberati}},
  \ and\ \bibinfo {author} {\bibfnamefont {D.}~\bibnamefont {Mattingly}},\
  }\href {\doibase 10.1016/j.aop.2005.06.004} {\bibfield  {journal} {\bibinfo
  {journal} {Annals Phys.}\ }\textbf {\bibinfo {volume} {321}},\ \bibinfo
  {pages} {150} (\bibinfo {year} {2006})},\ \Eprint
  {http://arxiv.org/abs/astro-ph/0505267} {arXiv:astro-ph/0505267 [astro-ph]}
  \BibitemShut {NoStop}%
%%CITATION = ASTRO-PH/0505267;%%
\bibitem [{\citenamefont {Tasson}(2014)}]{Tasson:2014dfa}%
  \BibitemOpen
  \bibfield  {author} {\bibinfo {author} {\bibfnamefont {J.~D.}\ \bibnamefont
  {Tasson}},\ }\href {\doibase 10.1088/0034-4885/77/6/062901} {\bibfield
  {journal} {\bibinfo  {journal} {Rept.\ Prog.\ Phys.}\ }\textbf {\bibinfo
  {volume} {77}},\ \bibinfo {pages} {062901} (\bibinfo {year} {2014})},\
  \Eprint {http://arxiv.org/abs/1403.7785} {arXiv:1403.7785 [hep-ph]}
  \BibitemShut {NoStop}%
%%CITATION = ARXIV:1403.7785;%%
\bibitem [{\citenamefont {Kosteleck\'y}\ and\ \citenamefont
  {Samuel}(1989)}]{Kostelecky:1988zi}%
  \BibitemOpen
  \bibfield  {author} {\bibinfo {author} {\bibfnamefont {V.~A.}\ \bibnamefont
  {Kosteleck\'y}}\ and\ \bibinfo {author} {\bibfnamefont {S.}~\bibnamefont
  {Samuel}},\ }\href {\doibase 10.1103/PhysRevD.39.683} {\bibfield  {journal}
  {\bibinfo  {journal} {Phys.\ Rev.\ D}\ }\textbf {\bibinfo {volume} {39}},\
  \bibinfo {pages} {683} (\bibinfo {year} {1989})}\BibitemShut {NoStop}%
%%CITATION = PHRVA,D39,683;%%
\bibitem [{\citenamefont {Kosteleck\'y}\ and\ \citenamefont
  {Potting}(1991)}]{Kostelecky:1991ak}%
  \BibitemOpen
  \bibfield  {author} {\bibinfo {author} {\bibfnamefont {V.~A.}\ \bibnamefont
  {Kosteleck\'y}}\ and\ \bibinfo {author} {\bibfnamefont {R.}~\bibnamefont
  {Potting}},\ }\href {\doibase 10.1016/0550-3213(91)90071-5} {\bibfield
  {journal} {\bibinfo  {journal} {Nucl.\ Phys.\ B}\ }\textbf {\bibinfo {volume}
  {359}},\ \bibinfo {pages} {545} (\bibinfo {year} {1991})}\BibitemShut
  {NoStop}%
%%CITATION = NUPHA,B359,545;%%
\bibitem [{\citenamefont {Mattingly}(2005)}]{Mattingly:2005re}%
  \BibitemOpen
  \bibfield  {author} {\bibinfo {author} {\bibfnamefont {D.}~\bibnamefont
  {Mattingly}},\ }\href {\doibase 10.12942/lrr-2005-5} {\bibfield  {journal}
  {\bibinfo  {journal} {Living Rev.\ Rel.}\ }\textbf {\bibinfo {volume} {8}},\
  \bibinfo {pages} {5} (\bibinfo {year} {2005})},\ \Eprint
  {http://arxiv.org/abs/gr-qc/0502097} {arXiv:gr-qc/0502097 [gr-qc]}
  \BibitemShut {NoStop}%
%%CITATION = GR-QC/0502097;%%
\bibitem [{\citenamefont {Kosteleck\'y}\ and\ \citenamefont
  {Russell}(2011)}]{Kostelecky:2008ts}%
  \BibitemOpen
  \bibfield  {author} {\bibinfo {author} {\bibfnamefont {V.~A.}\ \bibnamefont
  {Kosteleck\'y}}\ and\ \bibinfo {author} {\bibfnamefont {N.}~\bibnamefont
  {Russell}},\ }\href {\doibase 10.1103/RevModPhys.83.11} {\bibfield  {journal}
  {\bibinfo  {journal} {Rev.\ Mod.\ Phys.}\ }\textbf {\bibinfo {volume} {83}},\
  \bibinfo {pages} {11} (\bibinfo {year} {2011})},\ \Eprint
  {http://arxiv.org/abs/0801.0287} {arXiv:0801.0287 [hep-ph]} \BibitemShut
  {NoStop}%
%%CITATION = ARXIV:0801.0287;%%
\bibitem [{\citenamefont {Carroll}\ \emph {et~al.}(1990)\citenamefont
  {Carroll}, \citenamefont {Field},\ and\ \citenamefont
  {Jackiw}}]{Carroll:1989vb}%
  \BibitemOpen
  \bibfield  {author} {\bibinfo {author} {\bibfnamefont {S.~M.}\ \bibnamefont
  {Carroll}}, \bibinfo {author} {\bibfnamefont {G.~B.}\ \bibnamefont {Field}},
  \ and\ \bibinfo {author} {\bibfnamefont {R.}~\bibnamefont {Jackiw}},\ }\href
  {\doibase 10.1103/PhysRevD.41.1231} {\bibfield  {journal} {\bibinfo
  {journal} {Phys.\ Rev.\ D}\ }\textbf {\bibinfo {volume} {41}},\ \bibinfo
  {pages} {1231} (\bibinfo {year} {1990})}\BibitemShut {NoStop}%
%%CITATION = PHRVA,D41,1231;%%
\bibitem [{\citenamefont {Colladay}\ and\ \citenamefont
  {Kosteleck\'y}(1997)}]{Colladay:1996iz}%
  \BibitemOpen
  \bibfield  {author} {\bibinfo {author} {\bibfnamefont {D.}~\bibnamefont
  {Colladay}}\ and\ \bibinfo {author} {\bibfnamefont {V.~A.}\ \bibnamefont
  {Kosteleck\'y}},\ }\href {\doibase 10.1103/PhysRevD.55.6760} {\bibfield
  {journal} {\bibinfo  {journal} {Phys.\ Rev.\ D}\ }\textbf {\bibinfo {volume}
  {55}},\ \bibinfo {pages} {6760} (\bibinfo {year} {1997})},\ \Eprint
  {http://arxiv.org/abs/hep-ph/9703464} {arXiv:hep-ph/9703464 [hep-ph]}
  \BibitemShut {NoStop}%
%%CITATION = HEP-PH/9703464;%%
\bibitem [{\citenamefont {Colladay}\ and\ \citenamefont
  {Kosteleck\'y}(1998)}]{Colladay:1998fq}%
  \BibitemOpen
  \bibfield  {author} {\bibinfo {author} {\bibfnamefont {D.}~\bibnamefont
  {Colladay}}\ and\ \bibinfo {author} {\bibfnamefont {V.~A.}\ \bibnamefont
  {Kosteleck\'y}},\ }\href {\doibase 10.1103/PhysRevD.58.116002} {\bibfield
  {journal} {\bibinfo  {journal} {Phys.\ Rev.\ D}\ }\textbf {\bibinfo {volume}
  {58}},\ \bibinfo {pages} {116002} (\bibinfo {year} {1998})},\ \Eprint
  {http://arxiv.org/abs/hep-ph/9809521} {arXiv:hep-ph/9809521 [hep-ph]}
  \BibitemShut {NoStop}%
%%CITATION = HEP-PH/9809521;%%
\bibitem [{\citenamefont {Kosteleck\'y}(2004)}]{Kostelecky:2003fs}%
  \BibitemOpen
  \bibfield  {author} {\bibinfo {author} {\bibfnamefont {V.~A.}\ \bibnamefont
  {Kosteleck\'y}},\ }\href {\doibase 10.1103/PhysRevD.69.105009} {\bibfield
  {journal} {\bibinfo  {journal} {Phys.\ Rev.\ D}\ }\textbf {\bibinfo {volume}
  {69}},\ \bibinfo {pages} {105009} (\bibinfo {year} {2004})},\ \Eprint
  {http://arxiv.org/abs/hep-th/0312310} {arXiv:hep-th/0312310 [hep-th]}
  \BibitemShut {NoStop}%
%%CITATION = HEP-TH/0312310;%%
\bibitem [{\citenamefont {Hees}\ \emph {et~al.}(2016)\citenamefont {Hees},
  \citenamefont {Bailey}, \citenamefont {Bourgoin}, \citenamefont {Bars},
  \citenamefont {Guerlin},\ and\ \citenamefont
  {Le~Poncin-Lafitte}}]{Hees:2016lyw}%
  \BibitemOpen
  \bibfield  {author} {\bibinfo {author} {\bibfnamefont {A.}~\bibnamefont
  {Hees}}, \bibinfo {author} {\bibfnamefont {Q.~G.}\ \bibnamefont {Bailey}},
  \bibinfo {author} {\bibfnamefont {A.}~\bibnamefont {Bourgoin}}, \bibinfo
  {author} {\bibfnamefont {H.~P.-L.}\ \bibnamefont {Bars}}, \bibinfo {author}
  {\bibfnamefont {C.}~\bibnamefont {Guerlin}}, \ and\ \bibinfo {author}
  {\bibfnamefont {C.}~\bibnamefont {Le~Poncin-Lafitte}},\ }\href {\doibase
  10.3390/universe2040030} {\bibfield  {journal} {\bibinfo  {journal}
  {Universe}\ }\textbf {\bibinfo {volume} {2}},\ \bibinfo {pages} {30}
  (\bibinfo {year} {2016})},\ \Eprint {http://arxiv.org/abs/1610.04682}
  {arXiv:1610.04682 [gr-qc]} \BibitemShut {NoStop}%
%%CITATION = ARXIV:1610.04682;%%
\bibitem [{\citenamefont {Shao}\ and\ \citenamefont
  {Wex}(2016)}]{Shao:2016ezh}%
  \BibitemOpen
  \bibfield  {author} {\bibinfo {author} {\bibfnamefont {L.}~\bibnamefont
  {Shao}}\ and\ \bibinfo {author} {\bibfnamefont {N.}~\bibnamefont {Wex}},\
  }\href {\doibase 10.1007/s11433-016-0087-6} {\bibfield  {journal} {\bibinfo
  {journal} {Sci.\ China Phys.\ Mech.\ Astron.}\ }\textbf {\bibinfo {volume}
  {59}},\ \bibinfo {pages} {699501} (\bibinfo {year} {2016})},\ \Eprint
  {http://arxiv.org/abs/1604.03662} {arXiv:1604.03662 [gr-qc]} \BibitemShut
  {NoStop}%
%%CITATION = ARXIV:1604.03662;%%
\bibitem [{\citenamefont {Tasson}(2016)}]{Tasson:2016xib}%
  \BibitemOpen
  \bibfield  {author} {\bibinfo {author} {\bibfnamefont {J.~D.}\ \bibnamefont
  {Tasson}},\ }\href@noop {} {\bibfield  {journal} {\bibinfo  {journal}
  {Symmetry}\ }\textbf {\bibinfo {volume} {8}},\ \bibinfo {pages} {111}
  (\bibinfo {year} {2016})},\ \Eprint {http://arxiv.org/abs/1610.05357}
  {arXiv:1610.05357 [gr-qc]} \BibitemShut {NoStop}%
%%CITATION = ARXIV:1610.05357;%%
\bibitem [{\citenamefont {Will}(2014)}]{Will:2014kxa}%
  \BibitemOpen
  \bibfield  {author} {\bibinfo {author} {\bibfnamefont {C.~M.}\ \bibnamefont
  {Will}},\ }\href {\doibase 10.12942/lrr-2014-4} {\bibfield  {journal}
  {\bibinfo  {journal} {Living Rev.\ Rel.}\ }\textbf {\bibinfo {volume} {17}},\
  \bibinfo {pages} {4} (\bibinfo {year} {2014})},\ \Eprint
  {http://arxiv.org/abs/1403.7377} {arXiv:1403.7377 [gr-qc]} \BibitemShut
  {NoStop}%
%%CITATION = ARXIV:1403.7377;%%
\bibitem [{\citenamefont {Berti}\ \emph {et~al.}(2015)\citenamefont {Berti}
  \emph {et~al.}}]{Berti:2015itd}%
  \BibitemOpen
  \bibfield  {author} {\bibinfo {author} {\bibfnamefont {E.}~\bibnamefont
  {Berti}} \emph {et~al.},\ }\href {\doibase 10.1088/0264-9381/32/24/243001}
  {\bibfield  {journal} {\bibinfo  {journal} {Class.\ Quant.\ Grav.}\ }\textbf
  {\bibinfo {volume} {32}},\ \bibinfo {pages} {243001} (\bibinfo {year}
  {2015})},\ \Eprint {http://arxiv.org/abs/1501.07274} {arXiv:1501.07274
  [gr-qc]} \BibitemShut {NoStop}%
%%CITATION = ARXIV:1501.07274;%%
\bibitem [{\citenamefont {Amelino-Camelia}\ \emph {et~al.}(1998)\citenamefont
  {Amelino-Camelia}, \citenamefont {Ellis}, \citenamefont {Mavromatos},
  \citenamefont {Nanopoulos},\ and\ \citenamefont
  {Sarkar}}]{AmelinoCamelia:1997gz}%
  \BibitemOpen
  \bibfield  {author} {\bibinfo {author} {\bibfnamefont {G.}~\bibnamefont
  {Amelino-Camelia}}, \bibinfo {author} {\bibfnamefont {J.~R.}\ \bibnamefont
  {Ellis}}, \bibinfo {author} {\bibfnamefont {N.~E.}\ \bibnamefont
  {Mavromatos}}, \bibinfo {author} {\bibfnamefont {D.~V.}\ \bibnamefont
  {Nanopoulos}}, \ and\ \bibinfo {author} {\bibfnamefont {S.}~\bibnamefont
  {Sarkar}},\ }\href {\doibase 10.1038/31647} {\bibfield  {journal} {\bibinfo
  {journal} {Nature}\ }\textbf {\bibinfo {volume} {393}},\ \bibinfo {pages}
  {763} (\bibinfo {year} {1998})},\ \Eprint
  {http://arxiv.org/abs/astro-ph/9712103} {arXiv:astro-ph/9712103 [astro-ph]}
  \BibitemShut {NoStop}%
%%CITATION = ASTRO-PH/9712103;%%
\bibitem [{\citenamefont {Gambini}\ and\ \citenamefont
  {Pullin}(1999)}]{Gambini:1998it}%
  \BibitemOpen
  \bibfield  {author} {\bibinfo {author} {\bibfnamefont {R.}~\bibnamefont
  {Gambini}}\ and\ \bibinfo {author} {\bibfnamefont {J.}~\bibnamefont
  {Pullin}},\ }\href {\doibase 10.1103/PhysRevD.59.124021} {\bibfield
  {journal} {\bibinfo  {journal} {Phys.\ Rev.\ D}\ }\textbf {\bibinfo {volume}
  {59}},\ \bibinfo {pages} {124021} (\bibinfo {year} {1999})},\ \Eprint
  {http://arxiv.org/abs/gr-qc/9809038} {arXiv:gr-qc/9809038 [gr-qc]}
  \BibitemShut {NoStop}%
%%CITATION = GR-QC/9809038;%%
\bibitem [{\citenamefont {Amelino-Camelia}(2013)}]{AmelinoCamelia:2008qg}%
  \BibitemOpen
  \bibfield  {author} {\bibinfo {author} {\bibfnamefont {G.}~\bibnamefont
  {Amelino-Camelia}},\ }\href {\doibase 10.12942/lrr-2013-5} {\bibfield
  {journal} {\bibinfo  {journal} {Living Rev.\ Rel.}\ }\textbf {\bibinfo
  {volume} {16}},\ \bibinfo {pages} {5} (\bibinfo {year} {2013})},\ \Eprint
  {http://arxiv.org/abs/0806.0339} {arXiv:0806.0339 [gr-qc]} \BibitemShut
  {NoStop}%
%%CITATION = ARXIV:0806.0339;%%
\bibitem [{\citenamefont {Will}(2018)}]{Will:2018bme}%
  \BibitemOpen
  \bibfield  {author} {\bibinfo {author} {\bibfnamefont {C.~M.}\ \bibnamefont
  {Will}},\ }\href@noop {} {\emph {\bibinfo {title} {{Theory and Experiment in
  Gravitational Physics}}}}\ (\bibinfo  {publisher} {Cambridge University
  Press},\ \bibinfo {year} {2018})\BibitemShut {NoStop}%
%%CITATION = INSPIRE-1700339;%%
\bibitem [{\citenamefont {Kosteleck\'y}\ and\ \citenamefont
  {Tasson}(2011)}]{Kostelecky:2010ze}%
  \BibitemOpen
  \bibfield  {author} {\bibinfo {author} {\bibfnamefont {V.~A.}\ \bibnamefont
  {Kosteleck\'y}}\ and\ \bibinfo {author} {\bibfnamefont {J.~D.}\ \bibnamefont
  {Tasson}},\ }\href {\doibase 10.1103/PhysRevD.83.016013} {\bibfield
  {journal} {\bibinfo  {journal} {Phys.\ Rev.\ D}\ }\textbf {\bibinfo {volume}
  {83}},\ \bibinfo {pages} {016013} (\bibinfo {year} {2011})},\ \Eprint
  {http://arxiv.org/abs/1006.4106} {arXiv:1006.4106 [gr-qc]} \BibitemShut
  {NoStop}%
%%CITATION = ARXIV:1006.4106;%%
\bibitem [{\citenamefont {Bluhm}\ and\ \citenamefont
  {Kosteleck\'y}(2005)}]{Bluhm:2004ep}%
  \BibitemOpen
  \bibfield  {author} {\bibinfo {author} {\bibfnamefont {R.}~\bibnamefont
  {Bluhm}}\ and\ \bibinfo {author} {\bibfnamefont {V.~A.}\ \bibnamefont
  {Kosteleck\'y}},\ }\href {\doibase 10.1103/PhysRevD.71.065008} {\bibfield
  {journal} {\bibinfo  {journal} {Phys.\ Rev.\ D}\ }\textbf {\bibinfo {volume}
  {71}},\ \bibinfo {pages} {065008} (\bibinfo {year} {2005})},\ \Eprint
  {http://arxiv.org/abs/hep-th/0412320} {arXiv:hep-th/0412320 [hep-th]}
  \BibitemShut {NoStop}%
%%CITATION = HEP-TH/0412320;%%
\bibitem [{\citenamefont {Battat}\ \emph {et~al.}(2007)\citenamefont {Battat},
  \citenamefont {Chandler},\ and\ \citenamefont {Stubbs}}]{Battat:2007uh}%
  \BibitemOpen
  \bibfield  {author} {\bibinfo {author} {\bibfnamefont {J.~B.~R.}\
  \bibnamefont {Battat}}, \bibinfo {author} {\bibfnamefont {J.~F.}\
  \bibnamefont {Chandler}}, \ and\ \bibinfo {author} {\bibfnamefont {C.~W.}\
  \bibnamefont {Stubbs}},\ }\href {\doibase 10.1103/PhysRevLett.99.241103}
  {\bibfield  {journal} {\bibinfo  {journal} {Phys.\ Rev.\ Lett.}\ }\textbf
  {\bibinfo {volume} {99}},\ \bibinfo {pages} {241103} (\bibinfo {year}
  {2007})},\ \Eprint {http://arxiv.org/abs/0710.0702} {arXiv:0710.0702 [gr-qc]}
  \BibitemShut {NoStop}%
%%CITATION = ARXIV:0710.0702;%%
\bibitem [{\citenamefont {Bourgoin}\ \emph {et~al.}(2017)\citenamefont
  {Bourgoin}, \citenamefont {Le~Poncin-Lafitte}, \citenamefont {Hees},
  \citenamefont {Bouquillon}, \citenamefont {Francou},\ and\ \citenamefont
  {Angonin}}]{Bourgoin:2017fpo}%
  \BibitemOpen
  \bibfield  {author} {\bibinfo {author} {\bibfnamefont {A.}~\bibnamefont
  {Bourgoin}}, \bibinfo {author} {\bibfnamefont {C.}~\bibnamefont
  {Le~Poncin-Lafitte}}, \bibinfo {author} {\bibfnamefont {A.}~\bibnamefont
  {Hees}}, \bibinfo {author} {\bibfnamefont {S.}~\bibnamefont {Bouquillon}},
  \bibinfo {author} {\bibfnamefont {G.}~\bibnamefont {Francou}}, \ and\
  \bibinfo {author} {\bibfnamefont {M.-C.}\ \bibnamefont {Angonin}},\ }\href
  {\doibase 10.1103/PhysRevLett.119.201102} {\bibfield  {journal} {\bibinfo
  {journal} {Phys.\ Rev.\ Lett.}\ }\textbf {\bibinfo {volume} {119}},\ \bibinfo
  {pages} {201102} (\bibinfo {year} {2017})},\ \Eprint
  {http://arxiv.org/abs/1706.06294} {arXiv:1706.06294 [gr-qc]} \BibitemShut
  {NoStop}%
%%CITATION = ARXIV:1706.06294;%%
\bibitem [{\citenamefont {Mueller}\ \emph {et~al.}(2008)\citenamefont
  {Mueller}, \citenamefont {Chiow}, \citenamefont {Herrmann}, \citenamefont
  {Chu},\ and\ \citenamefont {Chung}}]{Muller:2007es}%
  \BibitemOpen
  \bibfield  {author} {\bibinfo {author} {\bibfnamefont {H.}~\bibnamefont
  {Mueller}}, \bibinfo {author} {\bibfnamefont {S.-w.}\ \bibnamefont {Chiow}},
  \bibinfo {author} {\bibfnamefont {S.}~\bibnamefont {Herrmann}}, \bibinfo
  {author} {\bibfnamefont {S.}~\bibnamefont {Chu}}, \ and\ \bibinfo {author}
  {\bibfnamefont {K.-Y.}\ \bibnamefont {Chung}},\ }\href {\doibase
  10.1103/PhysRevLett.100.031101} {\bibfield  {journal} {\bibinfo  {journal}
  {Phys.\ Rev.\ Lett.}\ }\textbf {\bibinfo {volume} {100}},\ \bibinfo {pages}
  {031101} (\bibinfo {year} {2008})},\ \Eprint {http://arxiv.org/abs/0710.3768}
  {arXiv:0710.3768 [gr-qc]} \BibitemShut {NoStop}%
%%CITATION = ARXIV:0710.3768;%%
\bibitem [{\citenamefont {Kosteleck\'y}\ and\ \citenamefont
  {Tasson}(2015)}]{Kostelecky:2015dpa}%
  \BibitemOpen
  \bibfield  {author} {\bibinfo {author} {\bibfnamefont {V.~A.}\ \bibnamefont
  {Kosteleck\'y}}\ and\ \bibinfo {author} {\bibfnamefont {J.~D.}\ \bibnamefont
  {Tasson}},\ }\href {\doibase 10.1016/j.physletb.2015.08.060} {\bibfield
  {journal} {\bibinfo  {journal} {Phys.\ Lett.\ B}\ }\textbf {\bibinfo {volume}
  {749}},\ \bibinfo {pages} {551} (\bibinfo {year} {2015})},\ \Eprint
  {http://arxiv.org/abs/1508.07007} {arXiv:1508.07007 [gr-qc]} \BibitemShut
  {NoStop}%
%%CITATION = ARXIV:1508.07007;%%
\bibitem [{\citenamefont {Shao}(2014{\natexlab{a}})}]{Shao:2014oha}%
  \BibitemOpen
  \bibfield  {author} {\bibinfo {author} {\bibfnamefont {L.}~\bibnamefont
  {Shao}},\ }\href {\doibase 10.1103/PhysRevLett.112.111103} {\bibfield
  {journal} {\bibinfo  {journal} {Phys.\ Rev.\ Lett.}\ }\textbf {\bibinfo
  {volume} {112}},\ \bibinfo {pages} {111103} (\bibinfo {year}
  {2014}{\natexlab{a}})},\ \Eprint {http://arxiv.org/abs/1402.6452}
  {arXiv:1402.6452 [gr-qc]} \BibitemShut {NoStop}%
%%CITATION = ARXIV:1402.6452;%%
\bibitem [{\citenamefont {Shao}(2014{\natexlab{b}})}]{Shao:2014bfa}%
  \BibitemOpen
  \bibfield  {author} {\bibinfo {author} {\bibfnamefont {L.}~\bibnamefont
  {Shao}},\ }\href {\doibase 10.1103/PhysRevD.90.122009} {\bibfield  {journal}
  {\bibinfo  {journal} {Phys.\ Rev.\ D}\ }\textbf {\bibinfo {volume} {90}},\
  \bibinfo {pages} {122009} (\bibinfo {year} {2014}{\natexlab{b}})},\ \Eprint
  {http://arxiv.org/abs/1412.2320} {arXiv:1412.2320 [gr-qc]} \BibitemShut
  {NoStop}%
%%CITATION = ARXIV:1412.2320;%%
\bibitem [{\citenamefont {Jennings}\ \emph {et~al.}(2015)\citenamefont
  {Jennings}, \citenamefont {Tasson},\ and\ \citenamefont
  {Yang}}]{Jennings:2015vma}%
  \BibitemOpen
  \bibfield  {author} {\bibinfo {author} {\bibfnamefont {R.~J.}\ \bibnamefont
  {Jennings}}, \bibinfo {author} {\bibfnamefont {J.~D.}\ \bibnamefont
  {Tasson}}, \ and\ \bibinfo {author} {\bibfnamefont {S.}~\bibnamefont
  {Yang}},\ }\href {\doibase 10.1103/PhysRevD.92.125028} {\bibfield  {journal}
  {\bibinfo  {journal} {Phys.\ Rev.\ D}\ }\textbf {\bibinfo {volume} {92}},\
  \bibinfo {pages} {125028} (\bibinfo {year} {2015})},\ \Eprint
  {http://arxiv.org/abs/1510.03798} {arXiv:1510.03798 [gr-qc]} \BibitemShut
  {NoStop}%
%%CITATION = ARXIV:1510.03798;%%
\bibitem [{\citenamefont {Shao}\ and\ \citenamefont
  {Bailey}(2018)}]{Shao:2018vul}%
  \BibitemOpen
  \bibfield  {author} {\bibinfo {author} {\bibfnamefont {L.}~\bibnamefont
  {Shao}}\ and\ \bibinfo {author} {\bibfnamefont {Q.~G.}\ \bibnamefont
  {Bailey}},\ }\href {\doibase 10.1103/PhysRevD.98.084049} {\bibfield
  {journal} {\bibinfo  {journal} {Phys.\ Rev.\ D}\ }\textbf {\bibinfo {volume}
  {98}},\ \bibinfo {pages} {084049} (\bibinfo {year} {2018})},\ \Eprint
  {http://arxiv.org/abs/1810.06332} {arXiv:1810.06332 [gr-qc]} \BibitemShut
  {NoStop}%
%%CITATION = ARXIV:1810.06332;%%
\bibitem [{\citenamefont {Shao}\ and\ \citenamefont
  {Bailey}(2019)}]{Shao:2019cyt}%
  \BibitemOpen
  \bibfield  {author} {\bibinfo {author} {\bibfnamefont {L.}~\bibnamefont
  {Shao}}\ and\ \bibinfo {author} {\bibfnamefont {Q.~G.}\ \bibnamefont
  {Bailey}},\ }\href {\doibase 10.1103/PhysRevD.99.084017} {\bibfield
  {journal} {\bibinfo  {journal} {Phys.\ Rev.\ D}\ }\textbf {\bibinfo {volume}
  {99}},\ \bibinfo {pages} {084017} (\bibinfo {year} {2019})},\ \Eprint
  {http://arxiv.org/abs/1903.11760} {arXiv:1903.11760 [gr-qc]} \BibitemShut
  {NoStop}%
%%CITATION = ARXIV:1903.11760;%%
\bibitem [{\citenamefont {Shao}(2019)}]{Shao:2019tle}%
  \BibitemOpen
  \bibfield  {author} {\bibinfo {author} {\bibfnamefont {L.}~\bibnamefont
  {Shao}},\ }\href {\doibase 10.3390/sym11091098} {\bibfield  {journal}
  {\bibinfo  {journal} {Symmetry}\ }\textbf {\bibinfo {volume} {11}},\ \bibinfo
  {pages} {1098} (\bibinfo {year} {2019})},\ \Eprint
  {http://arxiv.org/abs/1908.10019} {arXiv:1908.10019 [hep-ph]} \BibitemShut
  {NoStop}%
%%CITATION = ARXIV:1908.10019;%%
\bibitem [{\citenamefont {Hees}\ \emph {et~al.}(2015)\citenamefont {Hees},
  \citenamefont {Bailey}, \citenamefont {Le~Poncin-Lafitte}, \citenamefont
  {Bourgoin}, \citenamefont {Rivoldini}, \citenamefont {Lamine}, \citenamefont
  {Meynadier}, \citenamefont {Guerlin},\ and\ \citenamefont
  {Wolf}}]{Hees:2015mga}%
  \BibitemOpen
  \bibfield  {author} {\bibinfo {author} {\bibfnamefont {A.}~\bibnamefont
  {Hees}}, \bibinfo {author} {\bibfnamefont {Q.~G.}\ \bibnamefont {Bailey}},
  \bibinfo {author} {\bibfnamefont {C.}~\bibnamefont {Le~Poncin-Lafitte}},
  \bibinfo {author} {\bibfnamefont {A.}~\bibnamefont {Bourgoin}}, \bibinfo
  {author} {\bibfnamefont {A.}~\bibnamefont {Rivoldini}}, \bibinfo {author}
  {\bibfnamefont {B.}~\bibnamefont {Lamine}}, \bibinfo {author} {\bibfnamefont
  {F.}~\bibnamefont {Meynadier}}, \bibinfo {author} {\bibfnamefont
  {C.}~\bibnamefont {Guerlin}}, \ and\ \bibinfo {author} {\bibfnamefont
  {P.}~\bibnamefont {Wolf}},\ }\href {\doibase 10.1103/PhysRevD.92.064049}
  {\bibfield  {journal} {\bibinfo  {journal} {Phys.\ Rev.\ D}\ }\textbf
  {\bibinfo {volume} {92}},\ \bibinfo {pages} {064049} (\bibinfo {year}
  {2015})},\ \Eprint {http://arxiv.org/abs/1508.03478} {arXiv:1508.03478
  [gr-qc]} \BibitemShut {NoStop}%
%%CITATION = ARXIV:1508.03478;%%
\bibitem [{\citenamefont {Bailey}\ \emph {et~al.}(2015)\citenamefont {Bailey},
  \citenamefont {Kosteleck\'y},\ and\ \citenamefont {Xu}}]{Bailey:2014bta}%
  \BibitemOpen
  \bibfield  {author} {\bibinfo {author} {\bibfnamefont {Q.~G.}\ \bibnamefont
  {Bailey}}, \bibinfo {author} {\bibfnamefont {V.~A.}\ \bibnamefont
  {Kosteleck\'y}}, \ and\ \bibinfo {author} {\bibfnamefont {R.}~\bibnamefont
  {Xu}},\ }\href {\doibase 10.1103/PhysRevD.91.022006} {\bibfield  {journal}
  {\bibinfo  {journal} {Phys.\ Rev.\ D}\ }\textbf {\bibinfo {volume} {91}},\
  \bibinfo {pages} {022006} (\bibinfo {year} {2015})},\ \Eprint
  {http://arxiv.org/abs/1410.6162} {arXiv:1410.6162 [gr-qc]} \BibitemShut
  {NoStop}%
%%CITATION = ARXIV:1410.6162;%%
\bibitem [{\citenamefont {Shao}\ \emph {et~al.}(2016)\citenamefont {Shao} \emph
  {et~al.}}]{Shao:2016cjk}%
  \BibitemOpen
  \bibfield  {author} {\bibinfo {author} {\bibfnamefont {C.-G.}\ \bibnamefont
  {Shao}} \emph {et~al.},\ }\href {\doibase 10.1103/PhysRevLett.117.071102}
  {\bibfield  {journal} {\bibinfo  {journal} {Phys.\ Rev.\ Lett.}\ }\textbf
  {\bibinfo {volume} {117}},\ \bibinfo {pages} {071102} (\bibinfo {year}
  {2016})},\ \Eprint {http://arxiv.org/abs/1607.06095} {arXiv:1607.06095
  [gr-qc]} \BibitemShut {NoStop}%
%%CITATION = ARXIV:1607.06095;%%
\bibitem [{\citenamefont {Kosteleck\'y}\ and\ \citenamefont
  {Mewes}(2017)}]{Kostelecky:2016uex}%
  \BibitemOpen
  \bibfield  {author} {\bibinfo {author} {\bibfnamefont {V.~A.}\ \bibnamefont
  {Kosteleck\'y}}\ and\ \bibinfo {author} {\bibfnamefont {M.}~\bibnamefont
  {Mewes}},\ }\href {\doibase 10.1016/j.physletb.2016.12.062} {\bibfield
  {journal} {\bibinfo  {journal} {Phys.\ Lett.\ B}\ }\textbf {\bibinfo {volume}
  {766}},\ \bibinfo {pages} {137} (\bibinfo {year} {2017})},\ \Eprint
  {http://arxiv.org/abs/1611.10313} {arXiv:1611.10313 [gr-qc]} \BibitemShut
  {NoStop}%
%%CITATION = ARXIV:1611.10313;%%
\bibitem [{\citenamefont {Shao}\ \emph {et~al.}(2019)\citenamefont {Shao},
  \citenamefont {Chen}, \citenamefont {Tan}, \citenamefont {Yang},
  \citenamefont {Luo}, \citenamefont {Tobar}, \citenamefont {Long},
  \citenamefont {Weisman},\ and\ \citenamefont {Kosteleck\'y}}]{Shao:2018lsx}%
  \BibitemOpen
  \bibfield  {author} {\bibinfo {author} {\bibfnamefont {C.-G.}\ \bibnamefont
  {Shao}}, \bibinfo {author} {\bibfnamefont {Y.-F.}\ \bibnamefont {Chen}},
  \bibinfo {author} {\bibfnamefont {Y.-J.}\ \bibnamefont {Tan}}, \bibinfo
  {author} {\bibfnamefont {S.-Q.}\ \bibnamefont {Yang}}, \bibinfo {author}
  {\bibfnamefont {J.}~\bibnamefont {Luo}}, \bibinfo {author} {\bibfnamefont
  {M.~E.}\ \bibnamefont {Tobar}}, \bibinfo {author} {\bibfnamefont {J.~C.}\
  \bibnamefont {Long}}, \bibinfo {author} {\bibfnamefont {E.}~\bibnamefont
  {Weisman}}, \ and\ \bibinfo {author} {\bibfnamefont {V.~A.}\ \bibnamefont
  {Kosteleck\'y}},\ }\href {\doibase 10.1103/PhysRevLett.122.011102} {\bibfield
   {journal} {\bibinfo  {journal} {{Phys.\ Rev.\ Lett.}}\ }\textbf {\bibinfo
  {volume} {122}},\ \bibinfo {pages} {011102} (\bibinfo {year} {2019})},\
  \Eprint {http://arxiv.org/abs/1812.11123} {arXiv:1812.11123 [gr-qc]}
  \BibitemShut {NoStop}%
%%CITATION = ARXIV:1812.11123;%%
\bibitem [{\citenamefont {Flowers}\ \emph {et~al.}(2017)\citenamefont
  {Flowers}, \citenamefont {Goodge},\ and\ \citenamefont
  {Tasson}}]{Flowers:2016ctv}%
  \BibitemOpen
  \bibfield  {author} {\bibinfo {author} {\bibfnamefont {N.~A.}\ \bibnamefont
  {Flowers}}, \bibinfo {author} {\bibfnamefont {C.}~\bibnamefont {Goodge}}, \
  and\ \bibinfo {author} {\bibfnamefont {J.~D.}\ \bibnamefont {Tasson}},\
  }\href {\doibase 10.1103/PhysRevLett.119.201101} {\bibfield  {journal}
  {\bibinfo  {journal} {Phys.\ Rev.\ Lett.}\ }\textbf {\bibinfo {volume}
  {119}},\ \bibinfo {pages} {201101} (\bibinfo {year} {2017})},\ \Eprint
  {http://arxiv.org/abs/1612.08495} {arXiv:1612.08495 [gr-qc]} \BibitemShut
  {NoStop}%
%%CITATION = ARXIV:1612.08495;%%
\bibitem [{\citenamefont {Kosteleck\'y}\ and\ \citenamefont
  {Mewes}(2016)}]{Kostelecky:2016kfm}%
  \BibitemOpen
  \bibfield  {author} {\bibinfo {author} {\bibfnamefont {V.~A.}\ \bibnamefont
  {Kosteleck\'y}}\ and\ \bibinfo {author} {\bibfnamefont {M.}~\bibnamefont
  {Mewes}},\ }\href {\doibase 10.1016/j.physletb.2016.04.040} {\bibfield
  {journal} {\bibinfo  {journal} {Phys.\ Lett.\ B}\ }\textbf {\bibinfo {volume}
  {757}},\ \bibinfo {pages} {510} (\bibinfo {year} {2016})},\ \Eprint
  {http://arxiv.org/abs/1602.04782} {arXiv:1602.04782 [gr-qc]} \BibitemShut
  {NoStop}%
%%CITATION = ARXIV:1602.04782;%%
\bibitem [{\citenamefont {Mewes}(2019)}]{Mewes:2019dhj}%
  \BibitemOpen
  \bibfield  {author} {\bibinfo {author} {\bibfnamefont {M.}~\bibnamefont
  {Mewes}},\ }\href {\doibase 10.1103/PhysRevD.99.104062} {\bibfield  {journal}
  {\bibinfo  {journal} {Phys.\ Rev.\ D}\ }\textbf {\bibinfo {volume} {99}},\
  \bibinfo {pages} {104062} (\bibinfo {year} {2019})},\ \Eprint
  {http://arxiv.org/abs/1905.00409} {arXiv:1905.00409 [gr-qc]} \BibitemShut
  {NoStop}%
%%CITATION = ARXIV:1905.00409;%%
\bibitem [{\citenamefont {Xu}\ \emph {et~al.}(2020)\citenamefont {Xu},
  \citenamefont {Zhao},\ and\ \citenamefont {Shao}}]{Xu:2019gua}%
  \BibitemOpen
  \bibfield  {author} {\bibinfo {author} {\bibfnamefont {R.}~\bibnamefont
  {Xu}}, \bibinfo {author} {\bibfnamefont {J.}~\bibnamefont {Zhao}}, \ and\
  \bibinfo {author} {\bibfnamefont {L.}~\bibnamefont {Shao}},\ }\href {\doibase
  10.1016/j.physletb.2020.135283} {\bibfield  {journal} {\bibinfo  {journal}
  {Phys.\ Lett.\ B}\ }\textbf {\bibinfo {volume} {803}},\ \bibinfo {pages}
  {135283} (\bibinfo {year} {2020})},\ \Eprint
  {http://arxiv.org/abs/1909.10372} {arXiv:1909.10372 [gr-qc]} \BibitemShut
  {NoStop}%
\bibitem [{\citenamefont {Bailey}\ and\ \citenamefont
  {Havert}(2017)}]{Bailey:2017lbo}%
  \BibitemOpen
  \bibfield  {author} {\bibinfo {author} {\bibfnamefont {Q.~G.}\ \bibnamefont
  {Bailey}}\ and\ \bibinfo {author} {\bibfnamefont {D.}~\bibnamefont
  {Havert}},\ }\href {\doibase 10.1103/PhysRevD.96.064035} {\bibfield
  {journal} {\bibinfo  {journal} {Phys.\ Rev.\ D}\ }\textbf {\bibinfo {volume}
  {96}},\ \bibinfo {pages} {064035} (\bibinfo {year} {2017})},\ \Eprint
  {http://arxiv.org/abs/1706.10157} {arXiv:1706.10157 [gr-qc]} \BibitemShut
  {NoStop}%
%%CITATION = ARXIV:1706.10157;%%
\bibitem [{\citenamefont {Bailey}(2016)}]{Bailey:2016ezm}%
  \BibitemOpen
  \bibfield  {author} {\bibinfo {author} {\bibfnamefont {Q.~G.}\ \bibnamefont
  {Bailey}},\ }\href {\doibase 10.1103/PhysRevD.94.065029} {\bibfield
  {journal} {\bibinfo  {journal} {Phys.\ Rev.\ D}\ }\textbf {\bibinfo {volume}
  {94}},\ \bibinfo {pages} {065029} (\bibinfo {year} {2016})},\ \Eprint
  {http://arxiv.org/abs/1608.00267} {arXiv:1608.00267 [gr-qc]} \BibitemShut
  {NoStop}%
%%CITATION = ARXIV:1608.00267;%%
\bibitem [{\citenamefont {Abbott}\ \emph {et~al.}(2016)\citenamefont {Abbott}
  \emph {et~al.}}]{Abbott:2016blz}%
  \BibitemOpen
  \bibfield  {author} {\bibinfo {author} {\bibfnamefont {B.~P.}\ \bibnamefont
  {Abbott}} \emph {et~al.} (\bibinfo {collaboration} {Virgo, LIGO
  Scientific}),\ }\href {\doibase 10.1103/PhysRevLett.116.061102} {\bibfield
  {journal} {\bibinfo  {journal} {Phys.\ Rev.\ Lett.}\ }\textbf {\bibinfo
  {volume} {116}},\ \bibinfo {pages} {061102} (\bibinfo {year} {2016})},\
  \Eprint {http://arxiv.org/abs/1602.03837} {arXiv:1602.03837 [gr-qc]}
  \BibitemShut {NoStop}%
%%CITATION = ARXIV:1602.03837;%%
\bibitem [{\citenamefont {Abbott}\ \emph
  {et~al.}(2019{\natexlab{a}})\citenamefont {Abbott} \emph
  {et~al.}}]{LIGOScientific:2018mvr}%
  \BibitemOpen
  \bibfield  {author} {\bibinfo {author} {\bibfnamefont {B.~P.}\ \bibnamefont
  {Abbott}} \emph {et~al.} (\bibinfo {collaboration} {LIGO Scientific,
  Virgo}),\ }\href {\doibase 10.1103/PhysRevX.9.031040} {\bibfield  {journal}
  {\bibinfo  {journal} {Phys.\ Rev.\ X}\ }\textbf {\bibinfo {volume} {9}},\
  \bibinfo {pages} {031040} (\bibinfo {year} {2019}{\natexlab{a}})},\ \Eprint
  {http://arxiv.org/abs/1811.12907} {arXiv:1811.12907 [astro-ph.HE]}
  \BibitemShut {NoStop}%
%%CITATION = ARXIV:1811.12907;%%
\bibitem [{\citenamefont {Will}(1998)}]{Will:1997bb}%
  \BibitemOpen
  \bibfield  {author} {\bibinfo {author} {\bibfnamefont {C.~M.}\ \bibnamefont
  {Will}},\ }\href {\doibase 10.1103/PhysRevD.57.2061} {\bibfield  {journal}
  {\bibinfo  {journal} {Phys.\ Rev.\ D}\ }\textbf {\bibinfo {volume} {57}},\
  \bibinfo {pages} {2061} (\bibinfo {year} {1998})},\ \Eprint
  {http://arxiv.org/abs/gr-qc/9709011} {arXiv:gr-qc/9709011 [gr-qc]}
  \BibitemShut {NoStop}%
%%CITATION = GR-QC/9709011;%%
\bibitem [{\citenamefont {Mirshekari}\ \emph {et~al.}(2012)\citenamefont
  {Mirshekari}, \citenamefont {Yunes},\ and\ \citenamefont
  {Will}}]{Mirshekari:2011yq}%
  \BibitemOpen
  \bibfield  {author} {\bibinfo {author} {\bibfnamefont {S.}~\bibnamefont
  {Mirshekari}}, \bibinfo {author} {\bibfnamefont {N.}~\bibnamefont {Yunes}}, \
  and\ \bibinfo {author} {\bibfnamefont {C.~M.}\ \bibnamefont {Will}},\ }\href
  {\doibase 10.1103/PhysRevD.85.024041} {\bibfield  {journal} {\bibinfo
  {journal} {Phys.\ Rev.\ D}\ }\textbf {\bibinfo {volume} {85}},\ \bibinfo
  {pages} {024041} (\bibinfo {year} {2012})},\ \Eprint
  {http://arxiv.org/abs/1110.2720} {arXiv:1110.2720 [gr-qc]} \BibitemShut
  {NoStop}%
%%CITATION = ARXIV:1110.2720;%%
\bibitem [{\citenamefont {Yunes}\ \emph {et~al.}(2016)\citenamefont {Yunes},
  \citenamefont {Yagi},\ and\ \citenamefont {Pretorius}}]{Yunes:2016jcc}%
  \BibitemOpen
  \bibfield  {author} {\bibinfo {author} {\bibfnamefont {N.}~\bibnamefont
  {Yunes}}, \bibinfo {author} {\bibfnamefont {K.}~\bibnamefont {Yagi}}, \ and\
  \bibinfo {author} {\bibfnamefont {F.}~\bibnamefont {Pretorius}},\ }\href
  {\doibase 10.1103/PhysRevD.94.084002} {\bibfield  {journal} {\bibinfo
  {journal} {Phys.\ Rev.\ D}\ }\textbf {\bibinfo {volume} {94}},\ \bibinfo
  {pages} {084002} (\bibinfo {year} {2016})},\ \Eprint
  {http://arxiv.org/abs/1603.08955} {arXiv:1603.08955 [gr-qc]} \BibitemShut
  {NoStop}%
%%CITATION = ARXIV:1603.08955;%%
\bibitem [{\citenamefont {Abbott}\ \emph
  {et~al.}(2017{\natexlab{a}})\citenamefont {Abbott} \emph
  {et~al.}}]{Abbott:2017vtc}%
  \BibitemOpen
  \bibfield  {author} {\bibinfo {author} {\bibfnamefont {B.~P.}\ \bibnamefont
  {Abbott}} \emph {et~al.} (\bibinfo {collaboration} {VIRGO, LIGO
  Scientific}),\ }\href {\doibase 10.1103/PhysRevLett.118.221101} {\bibfield
  {journal} {\bibinfo  {journal} {Phys.\ Rev.\ Lett.}\ }\textbf {\bibinfo
  {volume} {118}},\ \bibinfo {pages} {221101} (\bibinfo {year}
  {2017}{\natexlab{a}})},\ \Eprint {http://arxiv.org/abs/1706.01812}
  {arXiv:1706.01812 [gr-qc]} \BibitemShut {NoStop}%
%%CITATION = ARXIV:1706.01812;%%
\bibitem [{\citenamefont {Nishizawa}(2018)}]{Nishizawa:2017nef}%
  \BibitemOpen
  \bibfield  {author} {\bibinfo {author} {\bibfnamefont {A.}~\bibnamefont
  {Nishizawa}},\ }\href {\doibase 10.1103/PhysRevD.97.104037} {\bibfield
  {journal} {\bibinfo  {journal} {Phys.\ Rev.\ D}\ }\textbf {\bibinfo {volume}
  {97}},\ \bibinfo {pages} {104037} (\bibinfo {year} {2018})},\ \Eprint
  {http://arxiv.org/abs/1710.04825} {arXiv:1710.04825 [gr-qc]} \BibitemShut
  {NoStop}%
%%CITATION = ARXIV:1710.04825;%%
\bibitem [{\citenamefont {Arai}\ and\ \citenamefont
  {Nishizawa}(2018)}]{Arai:2017hxj}%
  \BibitemOpen
  \bibfield  {author} {\bibinfo {author} {\bibfnamefont {S.}~\bibnamefont
  {Arai}}\ and\ \bibinfo {author} {\bibfnamefont {A.}~\bibnamefont
  {Nishizawa}},\ }\href {\doibase 10.1103/PhysRevD.97.104038} {\bibfield
  {journal} {\bibinfo  {journal} {Phys.\ Rev.\ D}\ }\textbf {\bibinfo {volume}
  {97}},\ \bibinfo {pages} {104038} (\bibinfo {year} {2018})},\ \Eprint
  {http://arxiv.org/abs/1711.03776} {arXiv:1711.03776 [gr-qc]} \BibitemShut
  {NoStop}%
%%CITATION = ARXIV:1711.03776;%%
\bibitem [{\citenamefont {Wang}(2017)}]{Wang:2017igw}%
  \BibitemOpen
  \bibfield  {author} {\bibinfo {author} {\bibfnamefont {S.}~\bibnamefont
  {Wang}},\ }\href@noop {} {\  (\bibinfo {year} {2017})},\ \Eprint
  {http://arxiv.org/abs/1712.06072} {arXiv:1712.06072 [gr-qc]} \BibitemShut
  {NoStop}%
%%CITATION = ARXIV:1712.06072;%%
\bibitem [{\citenamefont {Qiao}\ \emph {et~al.}(2019)\citenamefont {Qiao},
  \citenamefont {Zhu}, \citenamefont {Zhao},\ and\ \citenamefont
  {Wang}}]{Qiao:2019wsh}%
  \BibitemOpen
  \bibfield  {author} {\bibinfo {author} {\bibfnamefont {J.}~\bibnamefont
  {Qiao}}, \bibinfo {author} {\bibfnamefont {T.}~\bibnamefont {Zhu}}, \bibinfo
  {author} {\bibfnamefont {W.}~\bibnamefont {Zhao}}, \ and\ \bibinfo {author}
  {\bibfnamefont {A.}~\bibnamefont {Wang}},\ }\href {\doibase
  10.1103/PhysRevD.100.124058} {\bibfield  {journal} {\bibinfo  {journal}
  {Phys.\ Rev.\ D}\ }\textbf {\bibinfo {volume} {100}},\ \bibinfo {pages}
  {124058} (\bibinfo {year} {2019})},\ \Eprint
  {http://arxiv.org/abs/1909.03815} {arXiv:1909.03815 [gr-qc]} \BibitemShut
  {NoStop}%
%%CITATION = ARXIV:1909.03815;%%
\bibitem [{\citenamefont {Zhao}\ \emph {et~al.}(2020)\citenamefont {Zhao},
  \citenamefont {Zhu}, \citenamefont {Qiao},\ and\ \citenamefont
  {Wang}}]{Zhao:2019xmm}%
  \BibitemOpen
  \bibfield  {author} {\bibinfo {author} {\bibfnamefont {W.}~\bibnamefont
  {Zhao}}, \bibinfo {author} {\bibfnamefont {T.}~\bibnamefont {Zhu}}, \bibinfo
  {author} {\bibfnamefont {J.}~\bibnamefont {Qiao}}, \ and\ \bibinfo {author}
  {\bibfnamefont {A.}~\bibnamefont {Wang}},\ }\href {\doibase
  10.1103/PhysRevD.101.024002} {\bibfield  {journal} {\bibinfo  {journal}
  {Phys.\ Rev.\ D}\ }\textbf {\bibinfo {volume} {101}},\ \bibinfo {pages}
  {024002} (\bibinfo {year} {2020})},\ \Eprint
  {http://arxiv.org/abs/1909.10887} {arXiv:1909.10887 [gr-qc]} \BibitemShut
  {NoStop}%
%%CITATION = ARXIV:1909.10887;%%
\bibitem [{\citenamefont {Wang}\ and\ \citenamefont
  {Zhao}(2020)}]{Wang:2020pgu}%
  \BibitemOpen
  \bibfield  {author} {\bibinfo {author} {\bibfnamefont {S.}~\bibnamefont
  {Wang}}\ and\ \bibinfo {author} {\bibfnamefont {Z.-C.}\ \bibnamefont
  {Zhao}},\ }\href@noop {} {\  (\bibinfo {year} {2020})},\ \Eprint
  {http://arxiv.org/abs/2002.00396} {arXiv:2002.00396 [gr-qc]} \BibitemShut
  {NoStop}%
%%CITATION = ARXIV:2002.00396;%%
\bibitem [{\citenamefont {Abbott}\ \emph
  {et~al.}(2019{\natexlab{b}})\citenamefont {Abbott} \emph
  {et~al.}}]{LIGOScientific:2019fpa}%
  \BibitemOpen
  \bibfield  {author} {\bibinfo {author} {\bibfnamefont {B.~P.}\ \bibnamefont
  {Abbott}} \emph {et~al.} (\bibinfo {collaboration} {LIGO Scientific,
  Virgo}),\ }\href {\doibase 10.1103/PhysRevD.100.104036} {\bibfield  {journal}
  {\bibinfo  {journal} {Phys.\ Rev.\ D}\ }\textbf {\bibinfo {volume} {100}},\
  \bibinfo {pages} {104036} (\bibinfo {year} {2019}{\natexlab{b}})},\ \Eprint
  {http://arxiv.org/abs/1903.04467} {arXiv:1903.04467 [gr-qc]} \BibitemShut
  {NoStop}%
%%CITATION = ARXIV:1903.04467;%%
\bibitem [{\citenamefont {K.~Nordtvedt}(1976)}]{K.Nordtvedt:1976zz}%
  \BibitemOpen
  \bibfield  {author} {\bibinfo {author} {\bibfnamefont {J.}~\bibnamefont
  {K.~Nordtvedt}},\ }\href {\doibase 10.1103/PhysRevD.14.1511} {\bibfield
  {journal} {\bibinfo  {journal} {Phys.\ Rev.\ D}\ }\textbf {\bibinfo {volume}
  {14}},\ \bibinfo {pages} {1511} (\bibinfo {year} {1976})}\BibitemShut
  {NoStop}%
%%CITATION = PHRVA,D14,1511;%%
\bibitem [{\citenamefont {Bailey}\ and\ \citenamefont
  {Kosteleck\'y}(2006)}]{Bailey:2006fd}%
  \BibitemOpen
  \bibfield  {author} {\bibinfo {author} {\bibfnamefont {Q.~G.}\ \bibnamefont
  {Bailey}}\ and\ \bibinfo {author} {\bibfnamefont {V.~A.}\ \bibnamefont
  {Kosteleck\'y}},\ }\href {\doibase 10.1103/PhysRevD.74.045001} {\bibfield
  {journal} {\bibinfo  {journal} {Phys.\ Rev.\ D}\ }\textbf {\bibinfo {volume}
  {74}},\ \bibinfo {pages} {045001} (\bibinfo {year} {2006})},\ \Eprint
  {http://arxiv.org/abs/gr-qc/0603030} {arXiv:gr-qc/0603030 [gr-qc]}
  \BibitemShut {NoStop}%
%%CITATION = GR-QC/0603030;%%
\bibitem [{\citenamefont {Abbott}\ \emph
  {et~al.}(2017{\natexlab{b}})\citenamefont {Abbott} \emph
  {et~al.}}]{TheLIGOScientific:2017qsa}%
  \BibitemOpen
  \bibfield  {author} {\bibinfo {author} {\bibfnamefont {B.}~\bibnamefont
  {Abbott}} \emph {et~al.} (\bibinfo {collaboration} {Virgo, LIGO
  Scientific}),\ }\href {\doibase 10.1103/PhysRevLett.119.161101} {\bibfield
  {journal} {\bibinfo  {journal} {Phys.\ Rev.\ Lett.}\ }\textbf {\bibinfo
  {volume} {119}},\ \bibinfo {pages} {161101} (\bibinfo {year}
  {2017}{\natexlab{b}})},\ \Eprint {http://arxiv.org/abs/1710.05832}
  {arXiv:1710.05832 [gr-qc]} \BibitemShut {NoStop}%
%%CITATION = ARXIV:1710.05832;%%
\bibitem [{\citenamefont {Tasson}(2019)}]{Tasson:2019kuw}%
  \BibitemOpen
  \bibfield  {author} {\bibinfo {author} {\bibfnamefont {J.~D.}\ \bibnamefont
  {Tasson}},\ }in\ \href@noop {} {\emph {\bibinfo {booktitle} {{8th Meeting on
  CPT and Lorentz Symmetry (CPT'19) Bloomington, Indiana, USA, May 12-16,
  2019}}}}\ (\bibinfo {year} {2019})\ \Eprint {http://arxiv.org/abs/1907.08106}
  {arXiv:1907.08106 [hep-ph]} \BibitemShut {NoStop}%
%%CITATION = ARXIV:1907.08106;%%
\bibitem [{\citenamefont {Abbott}\ \emph
  {et~al.}(2019{\natexlab{c}})\citenamefont {Abbott} \emph
  {et~al.}}]{Abbott:2018lct}%
  \BibitemOpen
  \bibfield  {author} {\bibinfo {author} {\bibfnamefont {B.~P.}\ \bibnamefont
  {Abbott}} \emph {et~al.} (\bibinfo {collaboration} {LIGO Scientific,
  Virgo}),\ }\href {\doibase 10.1103/PhysRevLett.123.011102} {\bibfield
  {journal} {\bibinfo  {journal} {Phys.\ Rev.\ Lett.}\ }\textbf {\bibinfo
  {volume} {123}},\ \bibinfo {pages} {011102} (\bibinfo {year}
  {2019}{\natexlab{c}})},\ \Eprint {http://arxiv.org/abs/1811.00364}
  {arXiv:1811.00364 [gr-qc]} \BibitemShut {NoStop}%
%%CITATION = ARXIV:1811.00364;%%
\bibitem [{\citenamefont {Misner}\ \emph {et~al.}(1973)\citenamefont {Misner},
  \citenamefont {Thorne},\ and\ \citenamefont {Wheeler}}]{Misner:1974qy}%
  \BibitemOpen
  \bibfield  {author} {\bibinfo {author} {\bibfnamefont {C.~W.}\ \bibnamefont
  {Misner}}, \bibinfo {author} {\bibfnamefont {K.~S.}\ \bibnamefont {Thorne}},
  \ and\ \bibinfo {author} {\bibfnamefont {J.~A.}\ \bibnamefont {Wheeler}},\
  }\href@noop {} {\emph {\bibinfo {title} {{Gravitation}}}}\ (\bibinfo
  {publisher} {W. H. Freeman},\ \bibinfo {address} {San Francisco},\ \bibinfo
  {year} {1973})\BibitemShut {NoStop}%
%%CITATION = INSPIRE-95654;%%
\bibitem [{\citenamefont {Kosteleck\'y}\ and\ \citenamefont
  {Mewes}(2018)}]{Kostelecky:2017zob}%
  \BibitemOpen
  \bibfield  {author} {\bibinfo {author} {\bibfnamefont {V.~A.}\ \bibnamefont
  {Kosteleck\'y}}\ and\ \bibinfo {author} {\bibfnamefont {M.}~\bibnamefont
  {Mewes}},\ }\href {\doibase 10.1016/j.physletb.2018.01.082} {\bibfield
  {journal} {\bibinfo  {journal} {Phys.\ Lett.\ B}\ }\textbf {\bibinfo {volume}
  {779}},\ \bibinfo {pages} {136} (\bibinfo {year} {2018})},\ \Eprint
  {http://arxiv.org/abs/1712.10268} {arXiv:1712.10268 [gr-qc]} \BibitemShut
  {NoStop}%
%%CITATION = ARXIV:1712.10268;%%
\bibitem [{\citenamefont {Kosteleck\'y}\ and\ \citenamefont
  {Mewes}(2002)}]{Kostelecky:2002hh}%
  \BibitemOpen
  \bibfield  {author} {\bibinfo {author} {\bibfnamefont {V.~A.}\ \bibnamefont
  {Kosteleck\'y}}\ and\ \bibinfo {author} {\bibfnamefont {M.}~\bibnamefont
  {Mewes}},\ }\href {\doibase 10.1103/PhysRevD.66.056005} {\bibfield  {journal}
  {\bibinfo  {journal} {Phys.\ Rev.\ D}\ }\textbf {\bibinfo {volume} {66}},\
  \bibinfo {pages} {056005} (\bibinfo {year} {2002})},\ \Eprint
  {http://arxiv.org/abs/hep-ph/0205211} {arXiv:hep-ph/0205211 [hep-ph]}
  \BibitemShut {NoStop}%
%%CITATION = HEP-PH/0205211;%%
\bibitem [{\citenamefont {Kosteleck\'y}\ and\ \citenamefont
  {Mewes}(2009)}]{Kostelecky:2009zp}%
  \BibitemOpen
  \bibfield  {author} {\bibinfo {author} {\bibfnamefont {V.~A.}\ \bibnamefont
  {Kosteleck\'y}}\ and\ \bibinfo {author} {\bibfnamefont {M.}~\bibnamefont
  {Mewes}},\ }\href {\doibase 10.1103/PhysRevD.80.015020} {\bibfield  {journal}
  {\bibinfo  {journal} {Phys.\ Rev.\ D}\ }\textbf {\bibinfo {volume} {80}},\
  \bibinfo {pages} {015020} (\bibinfo {year} {2009})},\ \Eprint
  {http://arxiv.org/abs/0905.0031} {arXiv:0905.0031 [hep-ph]} \BibitemShut
  {NoStop}%
%%CITATION = ARXIV:0905.0031;%%
\bibitem [{GWT({\natexlab{a}})}]{GWTC1:catalog}%
  \BibitemOpen
  \href@noop {} {}\bibinfo {howpublished}
  {\url{https://doi.org/10.7935/82H3-HH23}} {\natexlab{a}}\BibitemShut
  {NoStop}%
\bibitem [{GWT({\natexlab{b}})}]{GWTC1:PE}%
  \BibitemOpen
  \href@noop {} {}\bibinfo {howpublished}
  {\url{https://dcc.ligo.org/LIGO-P1800370/public}}
  {\natexlab{b}}\BibitemShut {NoStop}%
\bibitem [{\citenamefont {Abbott}\ \emph
  {et~al.}(2019{\natexlab{d}})\citenamefont {Abbott} \emph
  {et~al.}}]{Abbott:2018wiz}%
  \BibitemOpen
  \bibfield  {author} {\bibinfo {author} {\bibfnamefont {B.~P.}\ \bibnamefont
  {Abbott}} \emph {et~al.} (\bibinfo {collaboration} {LIGO Scientific,
  Virgo}),\ }\href {\doibase 10.1103/PhysRevX.9.011001} {\bibfield  {journal}
  {\bibinfo  {journal} {Phys.\ Rev.\ X}\ }\textbf {\bibinfo {volume} {9}},\
  \bibinfo {pages} {011001} (\bibinfo {year} {2019}{\natexlab{d}})},\ \Eprint
  {http://arxiv.org/abs/1805.11579} {arXiv:1805.11579 [gr-qc]} \BibitemShut
  {NoStop}%
%%CITATION = ARXIV:1805.11579;%%
\bibitem [{\citenamefont {Jacob}\ and\ \citenamefont
  {Piran}(2008)}]{Jacob:2008bw}%
  \BibitemOpen
  \bibfield  {author} {\bibinfo {author} {\bibfnamefont {U.}~\bibnamefont
  {Jacob}}\ and\ \bibinfo {author} {\bibfnamefont {T.}~\bibnamefont {Piran}},\
  }\href {\doibase 10.1088/1475-7516/2008/01/031} {\bibfield  {journal}
  {\bibinfo  {journal} {JCAP}\ }\textbf {\bibinfo {volume} {0801}},\ \bibinfo
  {pages} {031} (\bibinfo {year} {2008})},\ \Eprint
  {http://arxiv.org/abs/0712.2170} {arXiv:0712.2170 [astro-ph]} \BibitemShut
  {NoStop}%
%%CITATION = ARXIV:0712.2170;%%
\bibitem [{\citenamefont {Aghanim}\ \emph {et~al.}(2018)\citenamefont {Aghanim}
  \emph {et~al.}}]{Aghanim:2018eyx}%
  \BibitemOpen
  \bibfield  {author} {\bibinfo {author} {\bibfnamefont {N.}~\bibnamefont
  {Aghanim}} \emph {et~al.} (\bibinfo {collaboration} {Planck}),\ }\href@noop
  {} {\  (\bibinfo {year} {2018})},\ \Eprint {http://arxiv.org/abs/1807.06209}
  {arXiv:1807.06209 [astro-ph.CO]} \BibitemShut {NoStop}%
%%CITATION = ARXIV:1807.06209;%%
\bibitem [{\citenamefont {Boh\'e}\ \emph {et~al.}(2017)\citenamefont {Boh\'e}
  \emph {et~al.}}]{Bohe:2016gbl}%
  \BibitemOpen
  \bibfield  {author} {\bibinfo {author} {\bibfnamefont {A.}~\bibnamefont
  {Boh\'e}} \emph {et~al.},\ }\href {\doibase 10.1103/PhysRevD.95.044028}
  {\bibfield  {journal} {\bibinfo  {journal} {Phys.\ Rev.\ D}\ }\textbf
  {\bibinfo {volume} {95}},\ \bibinfo {pages} {044028} (\bibinfo {year}
  {2017})},\ \Eprint {http://arxiv.org/abs/1611.03703} {arXiv:1611.03703
  [gr-qc]} \BibitemShut {NoStop}%
%%CITATION = ARXIV:1611.03703;%%
\bibitem [{\citenamefont {Shao}\ and\ \citenamefont {Ma}(2011)}]{Shao:2011uc}%
  \BibitemOpen
  \bibfield  {author} {\bibinfo {author} {\bibfnamefont {L.}~\bibnamefont
  {Shao}}\ and\ \bibinfo {author} {\bibfnamefont {B.-Q.}\ \bibnamefont {Ma}},\
  }\href {\doibase 10.1103/PhysRevD.83.127702} {\bibfield  {journal} {\bibinfo
  {journal} {Phys.\ Rev.\ D}\ }\textbf {\bibinfo {volume} {D83}},\ \bibinfo
  {pages} {127702} (\bibinfo {year} {2011})},\ \Eprint
  {http://arxiv.org/abs/1104.4438} {arXiv:1104.4438 [astro-ph.HE]} \BibitemShut
  {NoStop}%
%%CITATION = ARXIV:1104.4438;%%
\bibitem [{\citenamefont {Abbott}\ \emph
  {et~al.}(2017{\natexlab{c}})\citenamefont {Abbott} \emph
  {et~al.}}]{Monitor:2017mdv}%
  \BibitemOpen
  \bibfield  {author} {\bibinfo {author} {\bibfnamefont {B.~P.}\ \bibnamefont
  {Abbott}} \emph {et~al.} (\bibinfo {collaboration} {Virgo, Fermi-GBM,
  INTEGRAL, LIGO Scientific}),\ }\href {\doibase 10.3847/2041-8213/aa920c}
  {\bibfield  {journal} {\bibinfo  {journal} {Astrophys.\ J.}\ }\textbf
  {\bibinfo {volume} {848}},\ \bibinfo {pages} {L13} (\bibinfo {year}
  {2017}{\natexlab{c}})},\ \Eprint {http://arxiv.org/abs/1710.05834}
  {arXiv:1710.05834 [astro-ph.HE]} \BibitemShut {NoStop}%
%%CITATION = ARXIV:1710.05834;%%
\bibitem [{\citenamefont {Boyle}\ \emph {et~al.}(2019)\citenamefont {Boyle}
  \emph {et~al.}}]{Boyle:2019kee}%
  \BibitemOpen
  \bibfield  {author} {\bibinfo {author} {\bibfnamefont {M.}~\bibnamefont
  {Boyle}} \emph {et~al.},\ }\href {\doibase 10.1088/1361-6382/ab34e2}
  {\bibfield  {journal} {\bibinfo  {journal} {Class.\ Quant.\ Grav.}\ }\textbf
  {\bibinfo {volume} {36}},\ \bibinfo {pages} {195006} (\bibinfo {year}
  {2019})},\ \Eprint {http://arxiv.org/abs/1904.04831} {arXiv:1904.04831
  [gr-qc]} \BibitemShut {NoStop}%
%%CITATION = ARXIV:1904.04831;%%
\bibitem [{\citenamefont {Abbott}\ \emph
  {et~al.}(2017{\natexlab{d}})\citenamefont {Abbott} \emph
  {et~al.}}]{Abbott:2016wiq}%
  \BibitemOpen
  \bibfield  {author} {\bibinfo {author} {\bibfnamefont {B.~P.}\ \bibnamefont
  {Abbott}} \emph {et~al.} (\bibinfo {collaboration} {LIGO Scientific,
  Virgo}),\ }\href {\doibase 10.1088/1361-6382/aa6854} {\bibfield  {journal}
  {\bibinfo  {journal} {Class.\ Quant.\ Grav.}\ }\textbf {\bibinfo {volume}
  {34}},\ \bibinfo {pages} {104002} (\bibinfo {year} {2017}{\natexlab{d}})},\
  \Eprint {http://arxiv.org/abs/1611.07531} {arXiv:1611.07531 [gr-qc]}
  \BibitemShut {NoStop}%
\bibitem [{\citenamefont {Hinderer}\ and\ \citenamefont
  {Flanagan}(2008)}]{Hinderer:2008dm}%
  \BibitemOpen
  \bibfield  {author} {\bibinfo {author} {\bibfnamefont {T.}~\bibnamefont
  {Hinderer}}\ and\ \bibinfo {author} {\bibfnamefont {E.~E.}\ \bibnamefont
  {Flanagan}},\ }\href {\doibase 10.1103/PhysRevD.78.064028} {\bibfield
  {journal} {\bibinfo  {journal} {Phys.\ Rev.\ D}\ }\textbf {\bibinfo {volume}
  {78}},\ \bibinfo {pages} {064028} (\bibinfo {year} {2008})},\ \Eprint
  {http://arxiv.org/abs/0805.3337} {arXiv:0805.3337 [gr-qc]} \BibitemShut
  {NoStop}%
%%CITATION = ARXIV:0805.3337;%%
\bibitem [{\citenamefont {Maggiore}(2007)}]{Maggiore:1900zz}%
  \BibitemOpen
  \bibfield  {author} {\bibinfo {author} {\bibfnamefont {M.}~\bibnamefont
  {Maggiore}},\ }\href {http://www.oup.com/uk/catalogue/?ci=9780198570745}
  {\emph {\bibinfo {title} {{Gravitational Waves. Vol. 1: Theory and
  Experiments}}}},\ Oxford Master Series in Physics\ (\bibinfo  {publisher}
  {Oxford University Press},\ \bibinfo {year} {2007})\BibitemShut {NoStop}%
%%CITATION = INSPIRE-768483;%%
\bibitem [{\citenamefont {Abbott}\ \emph {et~al.}(2018)\citenamefont {Abbott}
  \emph {et~al.}}]{Aasi:2013wya}%
  \BibitemOpen
  \bibfield  {author} {\bibinfo {author} {\bibfnamefont {B.~P.}\ \bibnamefont
  {Abbott}} \emph {et~al.} (\bibinfo {collaboration} {VIRGO, KAGRA, LIGO
  Scientific}),\ }\href {\doibase 10.1007/s41114-018-0012-9,
  10.1007/lrr-2016-1} {\bibfield  {journal} {\bibinfo  {journal} {Living Rev.\
  Rel.}\ }\textbf {\bibinfo {volume} {21}},\ \bibinfo {pages} {3} (\bibinfo
  {year} {2018})},\ \Eprint {http://arxiv.org/abs/1304.0670} {arXiv:1304.0670
  [gr-qc]} \BibitemShut {NoStop}%
%%CITATION = ARXIV:1304.0670;%%
\bibitem [{\citenamefont {Virtanen}\ \emph {et~al.}(2020)\citenamefont
  {Virtanen} \emph {et~al.}}]{Virtanen:2019joe}%
  \BibitemOpen
  \bibfield  {author} {\bibinfo {author} {\bibfnamefont {P.}~\bibnamefont
  {Virtanen}} \emph {et~al.},\ }\href {\doibase 10.1038/s41592-019-0686-2}
  {\bibfield  {journal} {\bibinfo  {journal} {Nature Meth.}\ }\textbf {\bibinfo
  {volume} {17}},\ \bibinfo {pages} {261} (\bibinfo {year} {2020})},\ \Eprint
  {http://arxiv.org/abs/1907.10121} {arXiv:1907.10121 [cs.MS]} \BibitemShut
  {NoStop}%
\bibitem [{\citenamefont {Berti}\ \emph
  {et~al.}(2018{\natexlab{a}})\citenamefont {Berti}, \citenamefont {Yagi},\
  and\ \citenamefont {Yunes}}]{Berti:2018cxi}%
  \BibitemOpen
  \bibfield  {author} {\bibinfo {author} {\bibfnamefont {E.}~\bibnamefont
  {Berti}}, \bibinfo {author} {\bibfnamefont {K.}~\bibnamefont {Yagi}}, \ and\
  \bibinfo {author} {\bibfnamefont {N.}~\bibnamefont {Yunes}},\ }\href
  {\doibase 10.1007/s10714-018-2362-8} {\bibfield  {journal} {\bibinfo
  {journal} {Gen.\ Rel.\ Grav.}\ }\textbf {\bibinfo {volume} {50}},\ \bibinfo
  {pages} {46} (\bibinfo {year} {2018}{\natexlab{a}})},\ \Eprint
  {http://arxiv.org/abs/1801.03208} {arXiv:1801.03208 [gr-qc]} \BibitemShut
  {NoStop}%
%%CITATION = ARXIV:1801.03208;%%
\bibitem [{\citenamefont {Berti}\ \emph
  {et~al.}(2018{\natexlab{b}})\citenamefont {Berti}, \citenamefont {Yagi},
  \citenamefont {Yang},\ and\ \citenamefont {Yunes}}]{Berti:2018vdi}%
  \BibitemOpen
  \bibfield  {author} {\bibinfo {author} {\bibfnamefont {E.}~\bibnamefont
  {Berti}}, \bibinfo {author} {\bibfnamefont {K.}~\bibnamefont {Yagi}},
  \bibinfo {author} {\bibfnamefont {H.}~\bibnamefont {Yang}}, \ and\ \bibinfo
  {author} {\bibfnamefont {N.}~\bibnamefont {Yunes}},\ }\href {\doibase
  10.1007/s10714-018-2372-6} {\bibfield  {journal} {\bibinfo  {journal} {Gen.\
  Rel.\ Grav.}\ }\textbf {\bibinfo {volume} {50}},\ \bibinfo {pages} {49}
  (\bibinfo {year} {2018}{\natexlab{b}})},\ \Eprint
  {http://arxiv.org/abs/1801.03587} {arXiv:1801.03587 [gr-qc]} \BibitemShut
  {NoStop}%
%%CITATION = ARXIV:1801.03587;%%
\bibitem [{O3:()}]{O3:gracedb}%
  \BibitemOpen
  \href@noop {} {}\bibinfo {howpublished}
  {\url{https://gracedb.ligo.org/superevents/public/O3/}}\BibitemShut {NoStop}%
\bibitem [{\citenamefont {Abbott}\ \emph {et~al.}(2020)\citenamefont {Abbott}
  \emph {et~al.}}]{Abbott:2020uma}%
  \BibitemOpen
  \bibfield  {author} {\bibinfo {author} {\bibfnamefont {B.}~\bibnamefont
  {Abbott}} \emph {et~al.} (\bibinfo {collaboration} {LIGO Scientific,
  Virgo}),\ }\href {\doibase 10.3847/2041-8213/ab75f5} {\bibfield  {journal}
  {\bibinfo  {journal} {Astrophys.\ J.\ Lett.}\ }\textbf {\bibinfo {volume}
  {892}},\ \bibinfo {pages} {L3} (\bibinfo {year} {2020})},\ \Eprint
  {http://arxiv.org/abs/2001.01761} {arXiv:2001.01761 [astro-ph.HE]}
  \BibitemShut {NoStop}%
\bibitem [{\citenamefont {Aso}\ \emph {et~al.}(2013)\citenamefont {Aso},
  \citenamefont {Michimura}, \citenamefont {Somiya}, \citenamefont {Ando},
  \citenamefont {Miyakawa}, \citenamefont {Sekiguchi}, \citenamefont
  {Tatsumi},\ and\ \citenamefont {Yamamoto}}]{Aso:2013eba}%
  \BibitemOpen
  \bibfield  {author} {\bibinfo {author} {\bibfnamefont {Y.}~\bibnamefont
  {Aso}}, \bibinfo {author} {\bibfnamefont {Y.}~\bibnamefont {Michimura}},
  \bibinfo {author} {\bibfnamefont {K.}~\bibnamefont {Somiya}}, \bibinfo
  {author} {\bibfnamefont {M.}~\bibnamefont {Ando}}, \bibinfo {author}
  {\bibfnamefont {O.}~\bibnamefont {Miyakawa}}, \bibinfo {author}
  {\bibfnamefont {T.}~\bibnamefont {Sekiguchi}}, \bibinfo {author}
  {\bibfnamefont {D.}~\bibnamefont {Tatsumi}}, \ and\ \bibinfo {author}
  {\bibfnamefont {H.}~\bibnamefont {Yamamoto}} (\bibinfo {collaboration}
  {KAGRA}),\ }\href {\doibase 10.1103/PhysRevD.88.043007} {\bibfield  {journal}
  {\bibinfo  {journal} {Phys.\ Rev.\ D}\ }\textbf {\bibinfo {volume} {88}},\
  \bibinfo {pages} {043007} (\bibinfo {year} {2013})},\ \Eprint
  {http://arxiv.org/abs/1306.6747} {arXiv:1306.6747 [gr-qc]} \BibitemShut
  {NoStop}%
%%CITATION = ARXIV:1306.6747;%%
\bibitem [{GWT({\natexlab{c}})}]{GWTC1:skymaps}%
  \BibitemOpen
  \href@noop {} {}\bibinfo {howpublished}
  {\url{https://dcc.ligo.org/LIGO-P1800381/public}}
  {\natexlab{c}}\BibitemShut {NoStop}%
\bibitem [{GWS()}]{GWSOC}%
  \BibitemOpen
  \href@noop {} {}\bibinfo {howpublished}
  {\url{https://www.gw-openscience.org}}\BibitemShut {NoStop}%
\bibitem [{pac()}]{package:skymap}%
  \BibitemOpen
  \href@noop {} {}\bibinfo {howpublished}
  {\url{https://pypi.org/project/ligo.skymap}}\BibitemShut {NoStop}%
\bibitem [{\citenamefont {Newman}\ and\ \citenamefont
  {Penrose}(1966)}]{Newman:1966ub}%
  \BibitemOpen
  \bibfield  {author} {\bibinfo {author} {\bibfnamefont {E.~T.}\ \bibnamefont
  {Newman}}\ and\ \bibinfo {author} {\bibfnamefont {R.}~\bibnamefont
  {Penrose}},\ }\href {\doibase 10.1063/1.1931221} {\bibfield  {journal}
  {\bibinfo  {journal} {J.\ Math.\ Phys.}\ }\textbf {\bibinfo {volume} {7}},\
  \bibinfo {pages} {863} (\bibinfo {year} {1966})}\BibitemShut {NoStop}%
%%CITATION = JMAPA,7,863;%%
\bibitem [{\citenamefont {Goldberg}\ \emph {et~al.}(1967)\citenamefont
  {Goldberg}, \citenamefont {MacFarlane}, \citenamefont {Newman}, \citenamefont
  {Rohrlich},\ and\ \citenamefont {Sudarshan}}]{Goldberg:1966uu}%
  \BibitemOpen
  \bibfield  {author} {\bibinfo {author} {\bibfnamefont {J.~N.}\ \bibnamefont
  {Goldberg}}, \bibinfo {author} {\bibfnamefont {A.~J.}\ \bibnamefont
  {MacFarlane}}, \bibinfo {author} {\bibfnamefont {E.~T.}\ \bibnamefont
  {Newman}}, \bibinfo {author} {\bibfnamefont {F.}~\bibnamefont {Rohrlich}}, \
  and\ \bibinfo {author} {\bibfnamefont {E.~C.~G.}\ \bibnamefont {Sudarshan}},\
  }\href {\doibase 10.1063/1.1705135} {\bibfield  {journal} {\bibinfo
  {journal} {J.\ Math.\ Phys.}\ }\textbf {\bibinfo {volume} {8}},\ \bibinfo
  {pages} {2155} (\bibinfo {year} {1967})}\BibitemShut {NoStop}%
%%CITATION = JMAPA,8,2155;%%
\end{thebibliography}%
%---------------------------------------------------------------------

\end{document}